\documentclass[preprint,prd,tightenlines,superscriptaddress]{revtex4-1}

\usepackage{bm,amsmath,amsfonts,amssymb,graphicx,xspace,placeins,multirow}

\usepackage{graphicx} % Include figure files
\usepackage{dcolumn}  % Align table columns on decimal point
\usepackage{colordvi}
\usepackage{color}
\usepackage{epstopdf}
\usepackage{amssymb}
\usepackage{url}
\graphicspath{{ps}}
\usepackage{hyperref}
\usepackage{tabularx}
\usepackage{lineno}

\newcommand{\benu}{\ensuremath{\Bzb \to D^{*+} e^{-} \nub_e}\xspace}
\newcommand{\bmunu}{\ensuremath{\Bzb \to D^{*+} \mu^{-} \nub_\mu}\xspace}
\newcommand{\btaunu}{\ensuremath{\Bzb \to D^{(*)+} \tau^{-} \nub_\tau}\xspace}
\newcommand{\bdslnu}{\ensuremath{\Bzb \to D^{*+} \ell^{-} \nub_l}\xspace}
\newcommand{\bpdslnu}{\ensuremath{\Bm \to D^{*0} \ell^{-} \nub_l}\xspace}
\newcommand{\bdlnu}{\ensuremath{\Bm \to D^{0} \ell^{-} \nub_l}\xspace}

\newcommand{\ifb}{\ensuremath{{\rm fb}^{-1}}\xspace}
\newcommand{\lumi}{\ensuremath{34.6~\ifb}\xspace}
\newcommand{\BR}{{\ensuremath{\cal B}}}
\newcommand{\pis}{{\ensuremath{\pi_s}}\xspace}
\newcommand{\cosby}{\ensuremath{\cos\theta_{BY}}}

\newcommand{\resBFel}{\ensuremath{\BR(\benu) = \left(4.59 \pm0.06_{\mathrm{stat}}\pm0.48_{\mathrm{syst}}\right) \%}\xspace}
\newcommand{\resBFmu}{\ensuremath{\BR(\bmunu) = \left(4.62 \pm 0.06_{\mathrm{stat}}\pm0.49_{\mathrm{syst}}\right) \%}\xspace}
\newcommand{\resBF}{\ensuremath{\BR(\bdslnu) = \left(4.60 \pm 0.05_{\mathrm{stat}}\pm0.17_{\mathrm{syst}} \pm 0.45_{\pi_s}\right) \%}\xspace}

\newcommand{\NBB}{\ensuremath{N_{B \bar B}=  \left(37.7 \pm 0.6 \right) \times 10^6}}

\input ./belle2sym.tex

\begin{document}
%\linenumbers

%place for definitions and newcommands
\def\belletwo {\it {Belle II}}

\vspace*{-3\baselineskip}
\resizebox{!}{3cm}{\includegraphics{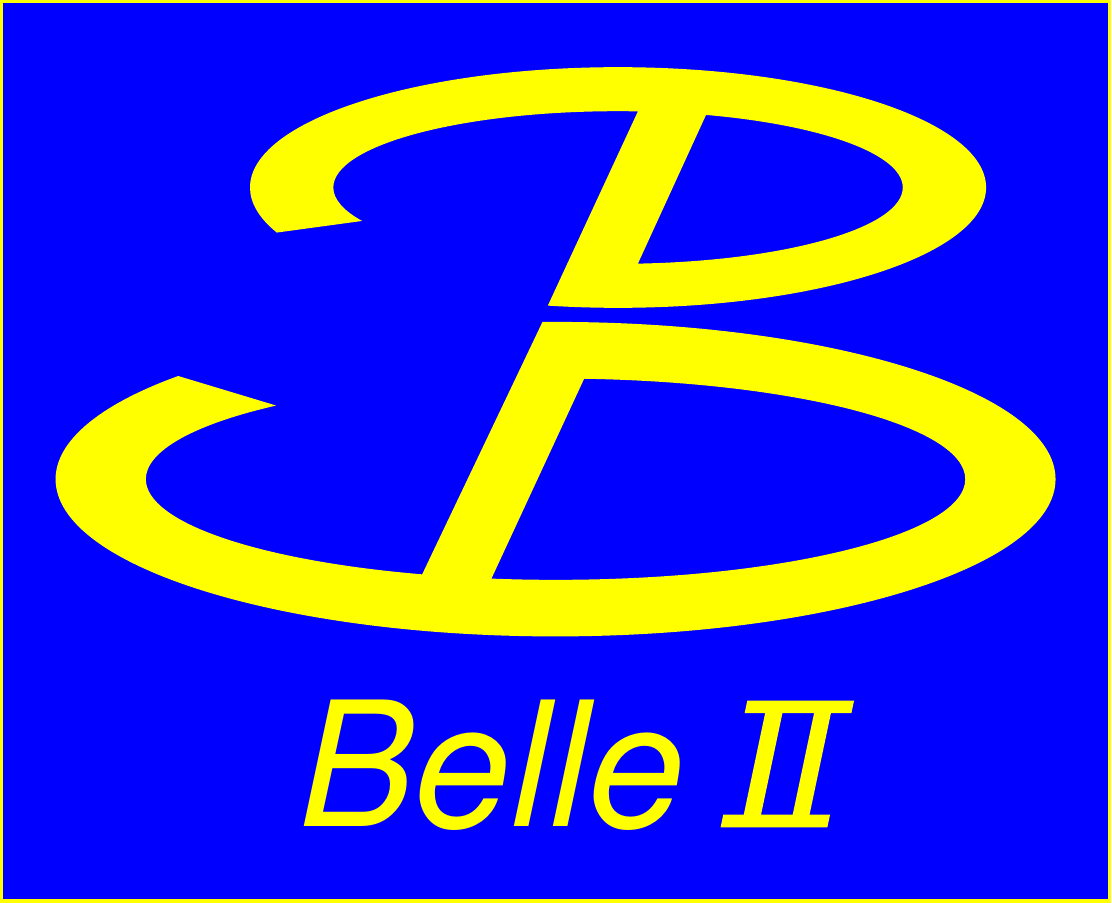}}

\vspace*{-5\baselineskip}
\begin{flushright}
BELLE2-CONF-PH-2020-008 \\
\today
\end{flushright}

\title { \quad\\[0.5cm] Studies of the semileptonic $\bar B^0\to D^{*+}\ell^-\bar\nu_\ell$ and $B^-\to D^{0}\ell^-\bar\nu_\ell$ decay processes with \mbox{34.6 fb$^{-1}$} of Belle~II data}

% \collaboration{The Belle II Collaboration}
%%% Paper:    (2020  conference papers)
%%% Journal:  (2020 conferences)
%%% ====================================================================
%%% Use \input{authors-conf2020} to insert this material into your latex file.
\newcommand{\instSinica}{Academia Sinica, Taipei 11529, Taiwan}
\newcommand{\instCPPM}{Aix Marseille Universit\'{e}, CNRS/IN2P3, CPPM, 13288 Marseille, France}
\newcommand{\instBeihang}{Beihang University, Beijing 100191, China}
\newcommand{\instBUAP}{Benemerita Universidad Autonoma de Puebla, Puebla 72570, Mexico}
\newcommand{\instBNL}{Brookhaven National Laboratory, Upton, New York 11973, U.S.A.}
\newcommand{\instBINP}{Budker Institute of Nuclear Physics SB RAS, Novosibirsk 630090, Russian Federation}
\newcommand{\instCMU}{Carnegie Mellon University, Pittsburgh, Pennsylvania 15213, U.S.A.}
\newcommand{\instCinvestavIPN}{Centro de Investigacion y de Estudios Avanzados del Instituto Politecnico Nacional, Mexico City 07360, Mexico}
\newcommand{\instPrague}{Faculty of Mathematics and Physics, Charles University, 121 16 Prague, Czech Republic}
\newcommand{\instChiangMai}{Chiang Mai University, Chiang Mai 50202, Thailand}
\newcommand{\instChiba}{Chiba University, Chiba 263-8522, Japan}
\newcommand{\instChonnam}{Chonnam National University, Gwangju 61186, South Korea}
\newcommand{\instConacyt}{Consejo Nacional de Ciencia y Tecnolog\'{\i}a, Mexico City 03940, Mexico}
\newcommand{\instDESY}{Deutsches Elektronen--Synchrotron, 22607 Hamburg, Germany}
\newcommand{\instDuke}{Duke University, Durham, North Carolina 27708, U.S.A.}
\newcommand{\instITAR}{Institute of Theoretical and Applied Research (ITAR), Duy Tan University, Hanoi 100000, Vietnam}
\newcommand{\instENEA}{ENEA Casaccia, I-00123 Roma, Italy}
\newcommand{\instEri}{Earthquake Research Institute, University of Tokyo, Tokyo 113-0032, Japan}
\newcommand{\instJuelich}{Forschungszentrum J\"{u}lich, 52425 J\"{u}lich, Germany}
\newcommand{\instFuJen}{Department of Physics, Fu Jen Catholic University, Taipei 24205, Taiwan}
\newcommand{\instFudan}{Key Laboratory of Nuclear Physics and Ion-beam Application (MOE) and Institute of Modern Physics, Fudan University, Shanghai 200443, China}
\newcommand{\instGoettingen}{II. Physikalisches Institut, Georg-August-Universit\"{a}t G\"{o}ttingen, 37073 G\"{o}ttingen, Germany}
\newcommand{\instGifu}{Gifu University, Gifu 501-1193, Japan}
\newcommand{\instSOKENDAI}{The Graduate University for Advanced Studies (SOKENDAI), Hayama 240-0193, Japan}
\newcommand{\instGyeongsang}{Gyeongsang National University, Jinju 52828, South Korea}
\newcommand{\instHanyang}{Department of Physics and Institute of Natural Sciences, Hanyang University, Seoul 04763, South Korea}
\newcommand{\instKEK}{High Energy Accelerator Research Organization (KEK), Tsukuba 305-0801, Japan}
\newcommand{\instJPARC}{J-PARC Branch, KEK Theory Center, High Energy Accelerator Research Organization (KEK), Tsukuba 305-0801, Japan}
\newcommand{\instHSE}{Higher School of Economics (HSE), Moscow 101000, Russian Federation}
\newcommand{\instIISER}{Indian Institute of Science Education and Research Mohali, SAS Nagar, 140306, India}
\newcommand{\instIITBhubaneswar}{Indian Institute of Technology Bhubaneswar, Satya Nagar 751007, India}
\newcommand{\instIITGuwahati}{Indian Institute of Technology Guwahati, Assam 781039, India}
\newcommand{\instIITHyderabad}{Indian Institute of Technology Hyderabad, Telangana 502285, India}
\newcommand{\instIITMadras}{Indian Institute of Technology Madras, Chennai 600036, India}
\newcommand{\instIndiana}{Indiana University, Bloomington, Indiana 47408, U.S.A.}
\newcommand{\instIHEPRussia}{Institute for High Energy Physics, Protvino 142281, Russian Federation}
\newcommand{\instHEPHYVienna}{Institute of High Energy Physics, Vienna 1050, Austria}
\newcommand{\instIHEPChina}{Institute of High Energy Physics, Chinese Academy of Sciences, Beijing 100049, China}
\newcommand{\instChennai}{Institute of Mathematical Sciences, Chennai 600113, India}
\newcommand{\instIPP}{Institute of Particle Physics (Canada), Victoria, British Columbia V8W 2Y2, Canada}
\newcommand{\instIOP}{Institute of Physics, Vietnam Academy of Science and Technology (VAST), Hanoi, Vietnam}
\newcommand{\instIFIC}{Instituto de Fisica Corpuscular, Paterna 46980, Spain}
\newcommand{\instFrascati}{INFN Laboratori Nazionali di Frascati, I-00044 Frascati, Italy}
\newcommand{\instNapoliINFN}{INFN Sezione di Napoli, I-80126 Napoli, Italy}
\newcommand{\instPadovaINFN}{INFN Sezione di Padova, I-35131 Padova, Italy}
\newcommand{\instPerugiaINFN}{INFN Sezione di Perugia, I-06123 Perugia, Italy}
\newcommand{\instPisaINFN}{INFN Sezione di Pisa, I-56127 Pisa, Italy}
\newcommand{\instRomaINFN}{INFN Sezione di Roma, I-00185 Roma, Italy}
\newcommand{\instRomaTreINFN}{INFN Sezione di Roma Tre, I-00146 Roma, Italy}
\newcommand{\instTorinoINFN}{INFN Sezione di Torino, I-10125 Torino, Italy}
\newcommand{\instTriesteINFN}{INFN Sezione di Trieste, I-34127 Trieste, Italy}
\newcommand{\instJAEA}{Advanced Science Research Center, Japan Atomic Energy Agency, Naka 319-1195, Japan}
\newcommand{\instMainz}{Johannes Gutenberg-Universit\"{a}t Mainz, Institut f\"{u}r Kernphysik, D-55099 Mainz, Germany}
\newcommand{\instGiessen}{Justus-Liebig-Universit\"{a}t Gie\ss{}en, 35392 Gie\ss{}en, Germany}
\newcommand{\instKarlsruhe}{Institut f\"{u}r Experimentelle Teilchenphysik, Karlsruher Institut f\"{u}r Technologie, 76131 Karlsruhe, Germany}
\newcommand{\instKennesaw}{Kennesaw State University, Kennesaw, Georgia 30144, U.S.A.}
\newcommand{\instKitasato}{Kitasato University, Sagamihara 252-0373, Japan}
\newcommand{\instKISTI}{Korea Institute of Science and Technology Information, Daejeon 34141, South Korea}
\newcommand{\instKorea}{Korea University, Seoul 02841, South Korea}
\newcommand{\instKSU}{Kyoto Sangyo University, Kyoto 603-8555, Japan}
\newcommand{\instKyotoU}{Kyoto University, Kyoto 606-8501, Japan}
\newcommand{\instKyungpook}{Kyungpook National University, Daegu 41566, South Korea}
\newcommand{\instLPI}{P.N. Lebedev Physical Institute of the Russian Academy of Sciences, Moscow 119991, Russian Federation}
\newcommand{\instLNNU}{Liaoning Normal University, Dalian 116029, China}
\newcommand{\instLMU}{Ludwig Maximilians University, 80539 Munich, Germany}
\newcommand{\instLuther}{Luther College, Decorah, Iowa 52101, U.S.A.}
\newcommand{\instMNITJaipur}{Malaviya National Institute of Technology Jaipur, Jaipur 302017, India}
\newcommand{\instMPP}{Max-Planck-Institut f\"{u}r Physik, 80805 M\"{u}nchen, Germany}
\newcommand{\instMPGHLL}{Semiconductor Laboratory of the Max Planck Society, 81739 M\"{u}nchen, Germany}
\newcommand{\instMcGill}{McGill University, Montr\'{e}al, Qu\'{e}bec, H3A 2T8, Canada}
\newcommand{\instMETU}{Middle East Technical University, 06531 Ankara, Turkey}
\newcommand{\instMEPhI}{Moscow Physical Engineering Institute, Moscow 115409, Russian Federation}
\newcommand{\instNagoya}{Graduate School of Science, Nagoya University, Nagoya 464-8602, Japan}
\newcommand{\instNagoyaKMI}{Kobayashi-Maskawa Institute, Nagoya University, Nagoya 464-8602, Japan}
\newcommand{\instNagoyaIAR}{Institute for Advanced Research, Nagoya University, Nagoya 464-8602, Japan}
\newcommand{\instNaraWu}{Nara Women's University, Nara 630-8506, Japan}
\newcommand{\instUNAM}{National Autonomous University of Mexico, Mexico City, Mexico}
\newcommand{\instNTUTaiwan}{Department of Physics, National Taiwan University, Taipei 10617, Taiwan}
\newcommand{\instNUUTaiwan}{National United University, Miao Li 36003, Taiwan}
\newcommand{\instKrakow}{H. Niewodniczanski Institute of Nuclear Physics, Krakow 31-342, Poland}
\newcommand{\instNiigata}{Niigata University, Niigata 950-2181, Japan}
\newcommand{\instNSU}{Novosibirsk State University, Novosibirsk 630090, Russian Federation}
\newcommand{\instOkinawa}{Okinawa Institute of Science and Technology, Okinawa 904-0495, Japan}
\newcommand{\instOsakaCity}{Osaka City University, Osaka 558-8585, Japan}
\newcommand{\instRCNP}{Research Center for Nuclear Physics, Osaka University, Osaka 567-0047, Japan}
\newcommand{\instPNNL}{Pacific Northwest National Laboratory, Richland, Washington 99352, U.S.A.}
\newcommand{\instPanjab}{Panjab University, Chandigarh 160014, India}
\newcommand{\instPeking}{Peking University, Beijing 100871, China}
\newcommand{\instPanjabPAU}{Punjab Agricultural University, Ludhiana 141004, India}
\newcommand{\instRIKENMSL}{Meson Science Laboratory, Cluster for Pioneering Research, RIKEN, Saitama 351-0198, Japan}
\newcommand{\instRIKEN}{Theoretical Research Division, Nishina Center, RIKEN, Saitama 351-0198, Japan}
\newcommand{\instXavier}{St. Francis Xavier University, Antigonish, Nova Scotia, B2G 2W5, Canada}
\newcommand{\instSeoul}{Seoul National University, Seoul 08826, South Korea}
\newcommand{\instShandong}{Shandong University, Jinan 250100, China}
\newcommand{\instSPU}{Showa Pharmaceutical University, Tokyo 194-8543, Japan}
\newcommand{\instSoochow}{Soochow University, Suzhou 215006, China}
\newcommand{\instSoongsil}{Soongsil University, Seoul 06978, South Korea}
\newcommand{\instLjubljanaJSI}{J. Stefan Institute, 1000 Ljubljana, Slovenia}
\newcommand{\instKyiv}{Taras Shevchenko National Univ. of Kiev, Kiev, Ukraine}
\newcommand{\instTata}{Tata Institute of Fundamental Research, Mumbai 400005, India}
\newcommand{\instTUM}{Department of Physics, Technische Universit\"{a}t M\"{u}nchen, 85748 Garching, Germany}
\newcommand{\instECUTUM}{Excellence Cluster Universe, Technische Universit\"{a}t M\"{u}nchen, 85748 Garching, Germany}
\newcommand{\instTelAviv}{Tel Aviv University, School of Physics and Astronomy, Tel Aviv, 69978, Israel}
\newcommand{\instToho}{Toho University, Funabashi 274-8510, Japan}
\newcommand{\instTohoku}{Department of Physics, Tohoku University, Sendai 980-8578, Japan}
\newcommand{\instTitech}{Tokyo Institute of Technology, Tokyo 152-8550, Japan}
\newcommand{\instTokyoMetropolitan}{Tokyo Metropolitan University, Tokyo 192-0397, Japan}
\newcommand{\instUAS}{Universidad Autonoma de Sinaloa, Sinaloa 80000, Mexico}
\newcommand{\instNapoliUNIV}{Dipartimento di Scienze Fisiche, Universit\`{a} di Napoli Federico II, I-80126 Napoli, Italy}
\newcommand{\instNapoliUNIVA}{Dipartimento di Agraria, Universit\`{a} di Napoli Federico II, I-80055 Portici (NA), Italy}
\newcommand{\instPadovaUNIV}{Dipartimento di Fisica e Astronomia, Universit\`{a} di Padova, I-35131 Padova, Italy}
\newcommand{\instPerugiaUNIV}{Dipartimento di Fisica, Universit\`{a} di Perugia, I-06123 Perugia, Italy}
\newcommand{\instPisaUNIV}{Dipartimento di Fisica, Universit\`{a} di Pisa, I-56127 Pisa, Italy}
\newcommand{\instRomaUNIV}{Universit\`{a} di Roma ``La Sapienza,'' I-00185 Roma, Italy}
\newcommand{\instRomaTreUNIV}{Dipartimento di Matematica e Fisica, Universit\`{a} di Roma Tre, I-00146 Roma, Italy}
\newcommand{\instTorinoUNIV}{Dipartimento di Fisica, Universit\`{a} di Torino, I-10125 Torino, Italy}
\newcommand{\instTriesteUNIV}{Dipartimento di Fisica, Universit\`{a} di Trieste, I-34127 Trieste, Italy}
\newcommand{\instMontreal}{Universit\'{e} de Montr\'{e}al, Physique des Particules, Montr\'{e}al, Qu\'{e}bec, H3C 3J7, Canada}
\newcommand{\instIJCLab}{Universit\'{e} Paris-Saclay, CNRS/IN2P3, IJCLab, 91405 Orsay, France}
\newcommand{\instIPHC}{Universit\'{e} de Strasbourg, CNRS, IPHC, UMR 7178, 67037 Strasbourg, France}
\newcommand{\instAdelaide}{Department of Physics, University of Adelaide, Adelaide, South Australia 5005, Australia}
\newcommand{\instBonn}{University of Bonn, 53115 Bonn, Germany}
\newcommand{\instUBC}{University of British Columbia, Vancouver, British Columbia, V6T 1Z1, Canada}
\newcommand{\instCincinnati}{University of Cincinnati, Cincinnati, Ohio 45221, U.S.A.}
\newcommand{\instFlorida}{University of Florida, Gainesville, Florida 32611, U.S.A.}
\newcommand{\instHamburg}{University of Hamburg, 20148 Hamburg, Germany}
\newcommand{\instHawaii}{University of Hawaii, Honolulu, Hawaii 96822, U.S.A.}
\newcommand{\instHeidelberg}{University of Heidelberg, 68131 Mannheim, Germany}
\newcommand{\instLjubljanaUniLJ}{Faculty of Mathematics and Physics, University of Ljubljana, 1000 Ljubljana, Slovenia}
\newcommand{\instLouisville}{University of Louisville, Louisville, Kentucky 40292, U.S.A.}
\newcommand{\instMalaya}{National Centre for Particle Physics, University Malaya, 50603 Kuala Lumpur, Malaysia}
\newcommand{\instLjubljanaUM}{University of Maribor, 2000 Maribor, Slovenia}
\newcommand{\instMelbourne}{School of Physics, University of Melbourne, Victoria 3010, Australia}
\newcommand{\instMississippi}{University of Mississippi, University, Mississippi 38677, U.S.A.}
\newcommand{\instUOM}{University of Miyazaki, Miyazaki 889-2192, Japan}
\newcommand{\instNovaGorica}{University of Nova Gorica, 5000 Nova Gorica, Slovenia}
\newcommand{\instPittsburgh}{University of Pittsburgh, Pittsburgh, Pennsylvania 15260, U.S.A.}
\newcommand{\instUSTC}{University of Science and Technology of China, Hefei 230026, China}
\newcommand{\instSAlabama}{University of South Alabama, Mobile, Alabama 36688, U.S.A.}
\newcommand{\instSCarolina}{University of South Carolina, Columbia, South Carolina 29208, U.S.A.}
\newcommand{\instSydney}{School of Physics, University of Sydney, New South Wales 2006, Australia}
\newcommand{\instTabuk}{Department of Physics, Faculty of Science, University of Tabuk, Tabuk 71451, Saudi Arabia}
\newcommand{\instUTokyo}{Department of Physics, University of Tokyo, Tokyo 113-0033, Japan}
\newcommand{\instIPMU}{Kavli Institute for the Physics and Mathematics of the Universe (WPI), University of Tokyo, Kashiwa 277-8583, Japan}
\newcommand{\instVictoria}{University of Victoria, Victoria, British Columbia, V8W 3P6, Canada}
\newcommand{\instVPI}{Virginia Polytechnic Institute and State University, Blacksburg, Virginia 24061, U.S.A.}
\newcommand{\instWayneState}{Wayne State University, Detroit, Michigan 48202, U.S.A.}
\newcommand{\instYamagata}{Yamagata University, Yamagata 990-8560, Japan}
\newcommand{\instYerevan}{Alikhanyan National Science Laboratory, Yerevan 0036, Armenia}
\newcommand{\instYonsei}{Yonsei University, Seoul 03722, South Korea}
%%%\affiliation{\instSinica}
\affiliation{\instCPPM}
\affiliation{\instBeihang}
%%%\affiliation{\instBUAP}
\affiliation{\instBNL}
\affiliation{\instBINP}
\affiliation{\instCMU}
\affiliation{\instCinvestavIPN}
\affiliation{\instPrague}
\affiliation{\instChiangMai}
\affiliation{\instChiba}
\affiliation{\instChonnam}
\affiliation{\instConacyt}
\affiliation{\instDESY}
\affiliation{\instDuke}
\affiliation{\instITAR}
%%%\affiliation{\instENEA}
\affiliation{\instEri}
\affiliation{\instJuelich}
\affiliation{\instFuJen}
\affiliation{\instFudan}
\affiliation{\instGoettingen}
\affiliation{\instGifu}
\affiliation{\instSOKENDAI}
\affiliation{\instGyeongsang}
\affiliation{\instHanyang}
\affiliation{\instKEK}
\affiliation{\instJPARC}
\affiliation{\instHSE}
\affiliation{\instIISER}
\affiliation{\instIITBhubaneswar}
\affiliation{\instIITGuwahati}
\affiliation{\instIITHyderabad}
\affiliation{\instIITMadras}
\affiliation{\instIndiana}
\affiliation{\instIHEPRussia}
\affiliation{\instHEPHYVienna}
\affiliation{\instIHEPChina}
%%%\affiliation{\instChennai}
\affiliation{\instIPP}
\affiliation{\instIOP}
\affiliation{\instIFIC}
\affiliation{\instFrascati}
\affiliation{\instNapoliINFN}
\affiliation{\instPadovaINFN}
\affiliation{\instPerugiaINFN}
\affiliation{\instPisaINFN}
\affiliation{\instRomaINFN}
\affiliation{\instRomaTreINFN}
\affiliation{\instTorinoINFN}
\affiliation{\instTriesteINFN}
\affiliation{\instJAEA}
\affiliation{\instMainz}
\affiliation{\instGiessen}
\affiliation{\instKarlsruhe}
%%%\affiliation{\instKennesaw}
\affiliation{\instKitasato}
\affiliation{\instKISTI}
\affiliation{\instKorea}
\affiliation{\instKSU}
%%%\affiliation{\instKyotoU}
\affiliation{\instKyungpook}
\affiliation{\instLPI}
\affiliation{\instLNNU}
\affiliation{\instLMU}
\affiliation{\instLuther}
\affiliation{\instMNITJaipur}
\affiliation{\instMPP}
\affiliation{\instMPGHLL}
\affiliation{\instMcGill}
%%%\affiliation{\instMETU}
\affiliation{\instMEPhI}
\affiliation{\instNagoya}
\affiliation{\instNagoyaKMI}
\affiliation{\instNagoyaIAR}
\affiliation{\instNaraWu}
%%%\affiliation{\instUNAM}
\affiliation{\instNTUTaiwan}
\affiliation{\instNUUTaiwan}
\affiliation{\instKrakow}
\affiliation{\instNiigata}
\affiliation{\instNSU}
\affiliation{\instOkinawa}
\affiliation{\instOsakaCity}
\affiliation{\instRCNP}
\affiliation{\instPNNL}
\affiliation{\instPanjab}
\affiliation{\instPeking}
\affiliation{\instPanjabPAU}
\affiliation{\instRIKENMSL}
%%%\affiliation{\instRIKEN}
%%%\affiliation{\instXavier}
\affiliation{\instSeoul}
%%%\affiliation{\instShandong}
\affiliation{\instSPU}
\affiliation{\instSoochow}
\affiliation{\instSoongsil}
\affiliation{\instLjubljanaJSI}
\affiliation{\instKyiv}
\affiliation{\instTata}
\affiliation{\instTUM}
%%%\affiliation{\instECUTUM}
\affiliation{\instTelAviv}
\affiliation{\instToho}
\affiliation{\instTohoku}
\affiliation{\instTitech}
\affiliation{\instTokyoMetropolitan}
\affiliation{\instUAS}
\affiliation{\instNapoliUNIV}
\affiliation{\instPadovaUNIV}
\affiliation{\instPerugiaUNIV}
\affiliation{\instPisaUNIV}
\affiliation{\instRomaUNIV}
\affiliation{\instRomaTreUNIV}
\affiliation{\instTorinoUNIV}
\affiliation{\instTriesteUNIV}
\affiliation{\instMontreal}
\affiliation{\instIJCLab}
\affiliation{\instIPHC}
\affiliation{\instAdelaide}
\affiliation{\instBonn}
\affiliation{\instUBC}
\affiliation{\instCincinnati}
\affiliation{\instFlorida}
%%%\affiliation{\instHamburg}
\affiliation{\instHawaii}
\affiliation{\instHeidelberg}
\affiliation{\instLjubljanaUniLJ}
\affiliation{\instLouisville}
\affiliation{\instMalaya}
\affiliation{\instLjubljanaUM}
\affiliation{\instMelbourne}
\affiliation{\instMississippi}
\affiliation{\instUOM}
%%%\affiliation{\instNovaGorica}
\affiliation{\instPittsburgh}
\affiliation{\instUSTC}
\affiliation{\instSAlabama}
\affiliation{\instSCarolina}
\affiliation{\instSydney}
%%%\affiliation{\instTabuk}
\affiliation{\instUTokyo}
\affiliation{\instIPMU}
\affiliation{\instVictoria}
\affiliation{\instVPI}
\affiliation{\instWayneState}
\affiliation{\instYamagata}
\affiliation{\instYerevan}
\affiliation{\instYonsei}
  \author{F.~Abudin{\'e}n}\affiliation{\instTriesteINFN} % 2250
  \author{I.~Adachi}\affiliation{\instKEK}\affiliation{\instSOKENDAI} % 2590
  \author{R.~Adak}\affiliation{\instFudan} % 6743
  \author{K.~Adamczyk}\affiliation{\instKrakow} % 2239
  \author{P.~Ahlburg}\affiliation{\instBonn} % 2367
  \author{J.~K.~Ahn}\affiliation{\instKorea} % 7423
  \author{H.~Aihara}\affiliation{\instUTokyo} % 2223
  \author{N.~Akopov}\affiliation{\instYerevan} % 9443
  \author{A.~Aloisio}\affiliation{\instNapoliUNIV}\affiliation{\instNapoliINFN} % 2194
  \author{F.~Ameli}\affiliation{\instRomaINFN} % 4683
% ...
% 2543
\collaboration{Belle II Collaboration}

\begin{abstract}
We report measurements of the \bdslnu and \bdlnu processes using \mbox{\lumi} of collision events recorded by the Belle~II experiment at the SuperKEKB asymmetric-energy $e^+ e^-$ collider. For the $B^-\to D^{0}\ell^-\bar\nu_\ell$ channel, we present first studies that isolate this decay from other semileptonic processes and backgrounds. We report a measurement of the \bdslnu branching fraction and obtain \resBF, in agreement with the world average. Here, the uncertainties are statistical, systematic, and related to slow pion reconstruction, respectively. The systematic uncertainties are limited by the statistics of auxiliary measurements and will improve in the future. We also report differential branching fractions in five bins of the hadronic recoil parameter $w$ for \bdslnu, unfolded to account for resolution and efficiency effects. 

\end{abstract}


\maketitle

{\renewcommand{\thefootnote}{\fnsymbol{footnote}}}
\setcounter{footnote}{0}

\section{Introduction}

Precision measurements of the decays of \bdslnu and \bdlnu ($\ell$ = $e$ or $\mu$) play an important role in the determination of the magnitude of the Cabibbo-Kobayashi-Maskawa matrix element $|V_{cb}|$ and are probes for the understanding of the hadronic dynamics of $B$ meson decays. These processes also constitute a source of background for measurements of charmless semileptonic decays and their understanding is important to study \btaunu. This motivates measurements of their branching fractions and kinematic distributions at Belle~II. The most precise measurements of $\BR(\bdslnu)$ and $\BR(\bdlnu)$ were obtained by the \babar~\cite{Aubert:2007rs,Aubert:2009ac} and Belle~\cite{Waheed:2018djm} collaborations. Since March 2019, the Belle~II experiment has been collecting $e^+ e^-$ collision events with the full detector and in this conference note studies, using an integrated luminosity of \lumi, are reported.

\section{The Belle~II detector and data sample}

The Belle~II detector~\cite{Abe:2010sj, ref:b2tip} operates at the SuperKEKB asymmetric-energy  electron-positron collider~\cite{superkekb}, located at the KEK laboratory in Tsukuba, Japan. The detector consists of several nested detector subsystems arranged around the beam pipe in a cylindrical geometry. The innermost subsystem is the vertex detector, which includes two layers of silicon pixel detectors and four outer layers of silicon strip detectors. Currently, the second pixel layer is installed in only a small part of the solid angle, while the remaining vertex detector layers are fully installed. Most of the tracking volume consists of a helium and ethane-based small-cell drift chamber. Outside the drift chamber, a Cherenkov-light imaging and time-of-propagation detector provides charged-particle identification in the barrel region. In the forward endcap, this function is provided by a proximity-focusing, ring-imaging Cherenkov detector with an aerogel radiator. Further out is an electromagnetic calorimeter, consisting of a barrel and two endcap sections made of CsI(Tl) crystals. A uniform 1.5~T magnetic field is provided by a superconducting solenoid situated outside the calorimeter. Multiple layers of scintillators and resistive plate chambers, located between the magnetic flux-return iron plates, constitute the $K_L$ and muon identification system.

The data used in this analysis were collected in 2019 and 2020 at a center-of-mass (CM) energy of 10.58~GeV, corresponding to the mass of the $\Upsilon$(4S) resonance. The energies of the electron and positron beams are $7\gev$ and $4\gev$, respectively, resulting in a boost of $\beta\gamma = 0.28$ of the CM frame relative to the lab frame. The number of $B$ meson pairs in the analyzed collision events is determined using event-shape variables to be \NBB.

Simulated Monte Carlo (MC) samples of signal events, with the subsequent decays $D^{*+}\to D^0\pi^+$ (for \bdslnu) and $D^0 \to K^- \pi^+$ (for both processes), are used to obtain the reconstruction efficiencies and signal kinematic distributions. These events are generated with EvtGen~\cite{Lange:2001uf}. Samples of background events are used to obtain kinematic distributions of the background. These include a sample of $e^+ e^-\to B\bar B$ with generic $B$ meson decays, generated with EvtGen, and correspond to an integrated luminosity of 100~\ifb and 200~\ifb for the \bdslnu and \bdlnu analyses, respectively.
A sample of continuum $e^+e^-\to q\bar q~ (q = u, d, s, c)$ is simulated with KKMC~\cite{Ward:2002qq} interfaced with PYTHIA~\cite{Sjostrand:2007gs}. All recorded collisions and simulated events are analyzed in the basf2~\cite{basf2} framework; a summary of the track reconstruction algorithms can be found in Ref.~\cite{Bertacchi:2020eez}.

\section{Event selection}

We reconstruct candidate events for both final states by reconstructing the $D^0\to K^- \pi^+$ decay and, for \bdslnu, the $D^{*+}\to D^0 \pi^+_s$ cascade. Here, $\pis$ indicates the soft pion originating from the $D^{*+}$ decay. Reconstruction of the charge-conjugate decays is implied.

Signal candidate reconstruction begins with the selection of charged-particle tracks. The distance of closest approach between each track and the interaction point is required to be less than 2~cm along the $z$ direction (parallel to the beams) and less than 0.5~cm in the transverse $r-\phi$ plane; the track's CM frame momentum must lie in the range $p_\ell^* \in [1.2,2.4]$~\gevc. The lepton candidate must also satisfy lepton-identification (lepton-ID) criteria based on information from all available detectors. A dedicated algorithm identifies photons from bremsstrahlung processes and corrects the momentum of reconstructed electron candidates if such can be identified. Given the high purity of the \bdslnu decay chain, application of kaon or pion identification criteria is deemed unnecessary and is thus not performed. For the \bdlnu decay, we apply loose kaon and pion identification criteria to increase the purity of the selected events. 

\subsection{\bdslnu Reconstruction}

From the \bdslnu selection, a vertex fit is applied to the $D^0$ candidate, constraining its $K^-\pi^+$ daughter tracks to originate from a common point. The invariant mass of the $D^0$ candidate is required to satisfy \mbox{$m_{K\pi} \in [1.85, 1.88]$~\gevcc} after the fit. The $D^{*+}\to D^0\pi^+_s$  candidate decay is also subjected to a vertex fit, after which the mass difference between the $D^*$ and $D^0$ candidates is required to satisfy $\Delta m \in [0.144,0.148]$~\gevcc. Continuum background is suppressed by requiring the momentum of the $D^*$ candidate in the CM frame to be less than 2.5~\gevc. Further continuum suppression is achieved by requiring $R_2 < 0.3$, where $R_2$ is the ratio of the second and zeroth Fox-Wolfram moments~\cite{Fox:1978vu}, calculated using all the tracks and photon candidates in the event. After applying all the selection criteria above, multiple \bdslnu candidates are found in only about 2\% of the events. For all candidates, we perform a vertex fit for the decay \bdslnu and, in events with multiple candidates per event, we select the candidate with the smallest value of the vertex-fit $\chi^2$. The signal efficiency after all selection criteria is $\epsilon = (21.3 \pm 2.2)\%$ for \benu and $\epsilon = (21.8 \pm 2.2)\%$ for \bmunu. These values are obtained from signal MC with lepton-ID efficiency corrections obtained from data-MC comparisons of reconstructed $J/\psi\to\ell^+\ell^-$, $e^+e^-\to \ell^+\ell^-$ and $e^+e^-\to e^+e^-\ell^+\ell^-$ decays. The quoted uncertainties are dominated by the uncertainties on the slow pion reconstruction efficiency. This uncertainty is estimated by studying slow pions from $B \to D^{*} \pi$ and $B \to D^* \rho$ decays, and will be reduced in the future. 

\subsection{\bdlnu Reconstruction}

To reduce the sizeable background of \bdslnu and \bpdslnu decays in the reconstructed \bdlnu candidates, an active veto is applied. This is done by combining charged and neutral soft pion candidates and photons to explicitly reconstruct the $D^{*+}\to D^0\pi^+_s$, $D^{*\, 0} \to D^{0\,} \pi^0$ and $D^{*\, 0} \to D^{0\,} \gamma$ decay cascades. Candidates using slow pions (charged or neutral) or photons are vetoed if a combination is found with $\Delta m \in [0.144,0.148]$~\gevcc or $\Delta m \in [0.141,0.145]$~\gevcc, respectively. To further control these backgrounds, a multivariate classifier in the form of a deep neural network is trained. Its input layer consists of the four-momenta of the final state particles and variables characterizing cluster properties in the electromagnetic calorimeter. The latter can be used to identify further neutral soft pions and photons from $D^{*\, 0} \to D^{0\,} \pi^0$ and  $D^{*\, 0} \to D^{0\,} \gamma$ decays, which were missed in the explicit reconstruction. The most important distinguishing input feature to veto \bdslnu events are the $D^0$ and lepton momenta. Finally, we demand that the invariant mass of the $D^0 \ell$ system is smaller than 3.15 GeV$/c^2$ and the momentum of the candidate lepton in the laboratory frame is below 3 GeV$/c$. No best candidate selection is carried out and all candidate events are analyzed. 

\section{Signal and Background Separation}

For each candidate, we calculate the angle between the $Y = D^{*+} \ell$ or $Y = D^0 \ell$ system and the $B$ meson in the center-of-mass frame of the collision. It can be determined using the reconstructed momenta and energies via
\begin{equation}
\cosby = {2\, E_B^* E_Y^* - m_B^2 - m_Y^2 \over 2 |p_B^*||p_Y^*|},
\end{equation}
where $E_Y^*$, $|p_Y^*|$, and $m_Y$ are the CM energy, momentum, and invariant mass, respectively, of the $D^{*+} \ell$ or $D^0 \ell$ system, $m_B$ is the nominal $B$ mass~\cite{Zyla:2020},
and $E_B^*$, $|p_B^*|$ are the CM energy and momentum, respectively, of the $B$; the CM is inferred from the beam four-momenta.  For correctly reconstructed \bdlnu and \bdslnu\ candidates with perfect detector resolution and correct values of $E_B^*$ and $p_B^*$, the value of \cosby\ ranges within the geometric range of $[-1,1]$. Due to the finite beam-energy spread, final-state radiation, and detector resolution, the \cosby\ distributions of signal events are smeared beyond this range, but retain an excellent sensitivity to separate signal from background processes. 

\subsection{Signal Yield Determination}

We determine the \bdslnu and \bdlnu signal event yields by carrying out a binned maximum-likelihood fit to the \cosby\ distribution. The probability density functions (PDFs) used in this fit are determined from simulated samples. We apply momentum- and polar-angle-dependent corrections to the lepton-identification efficiencies of leptons and hadrons. For leptons, corrections of the order of a few percent are obtained from $J/\psi\to \ell^+\ell^- ~(\ell = e, \mu)$ decays. Corrections for hadrons misidentified as leptons are obtained from samples of reconstructed $K_S \to \pi^+ \pi^-$ decays. The \bdlnu fit uses four components, for signal, $D^*$ background from \bdslnu and \bpdslnu, background from other $B \overline B$ processes, and continuum processes. The \bdslnu fit uses three components for signal, background from $B$ mesons, and continuum processes. 

Figure~\ref{fig:post_fit_results} shows the fitted \cosby\ distributions for \bdlnu and \bdslnu. The fitted distribution describe the measured spectra well. The selected \bdlnu candidates have a sizeable contamination from \bdslnu processes, but the signal can be clearly isolated. In total, we find $6186 \pm 234$ and $5800 \pm 231$ \bdlnu candidates in the electron and muon channels, respectively. The \bdslnu sample is much cleaner, in contrast, and we find $9583 \pm 134$ and  $9860 \pm 132$ signal events in the electron and muon channels, respectively. 

\begin{figure}
\begin{center}
\begin{tabular}{cccc}
\includegraphics[width=0.5\columnwidth]{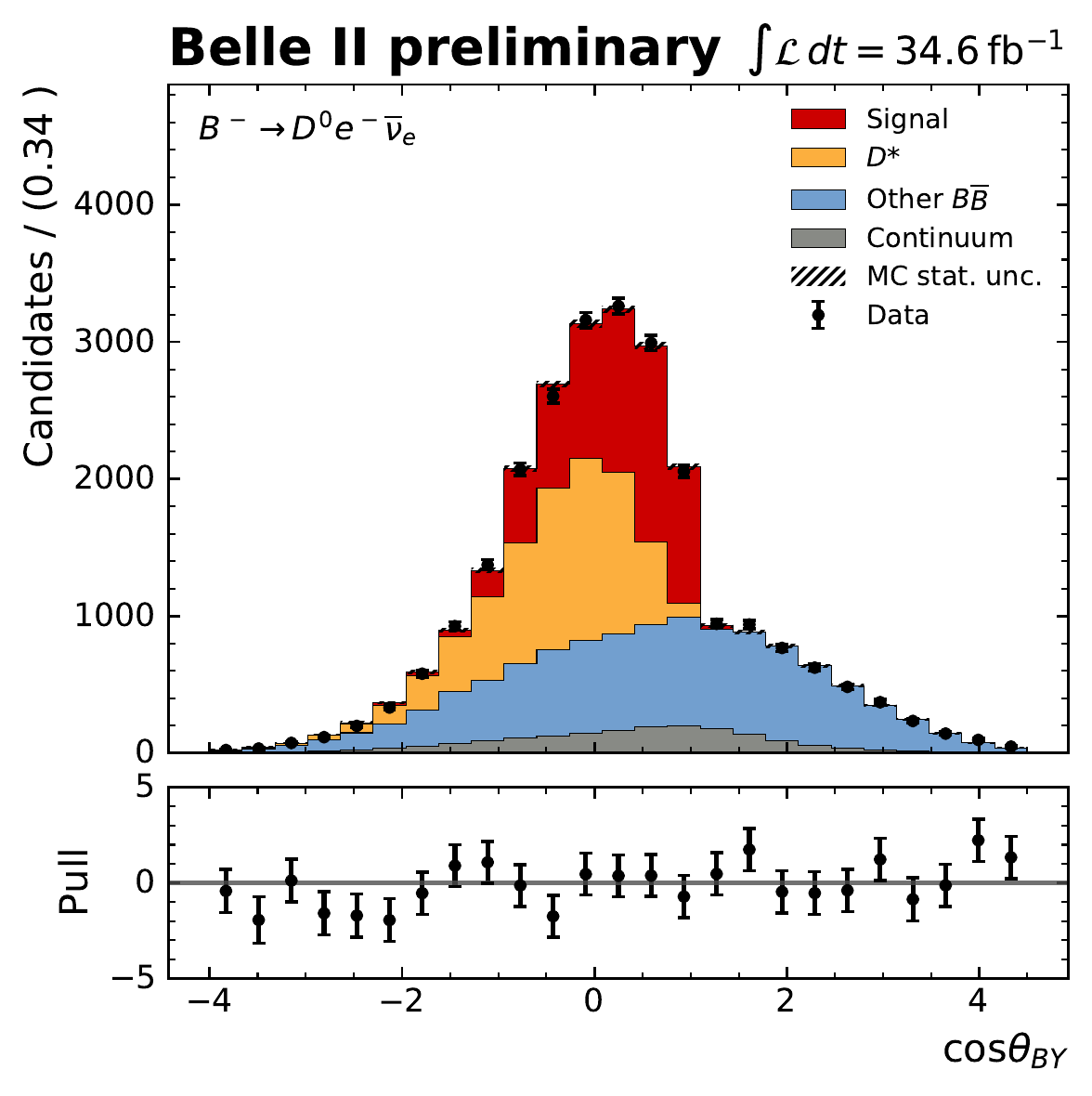} &
\includegraphics[width=0.5\columnwidth]{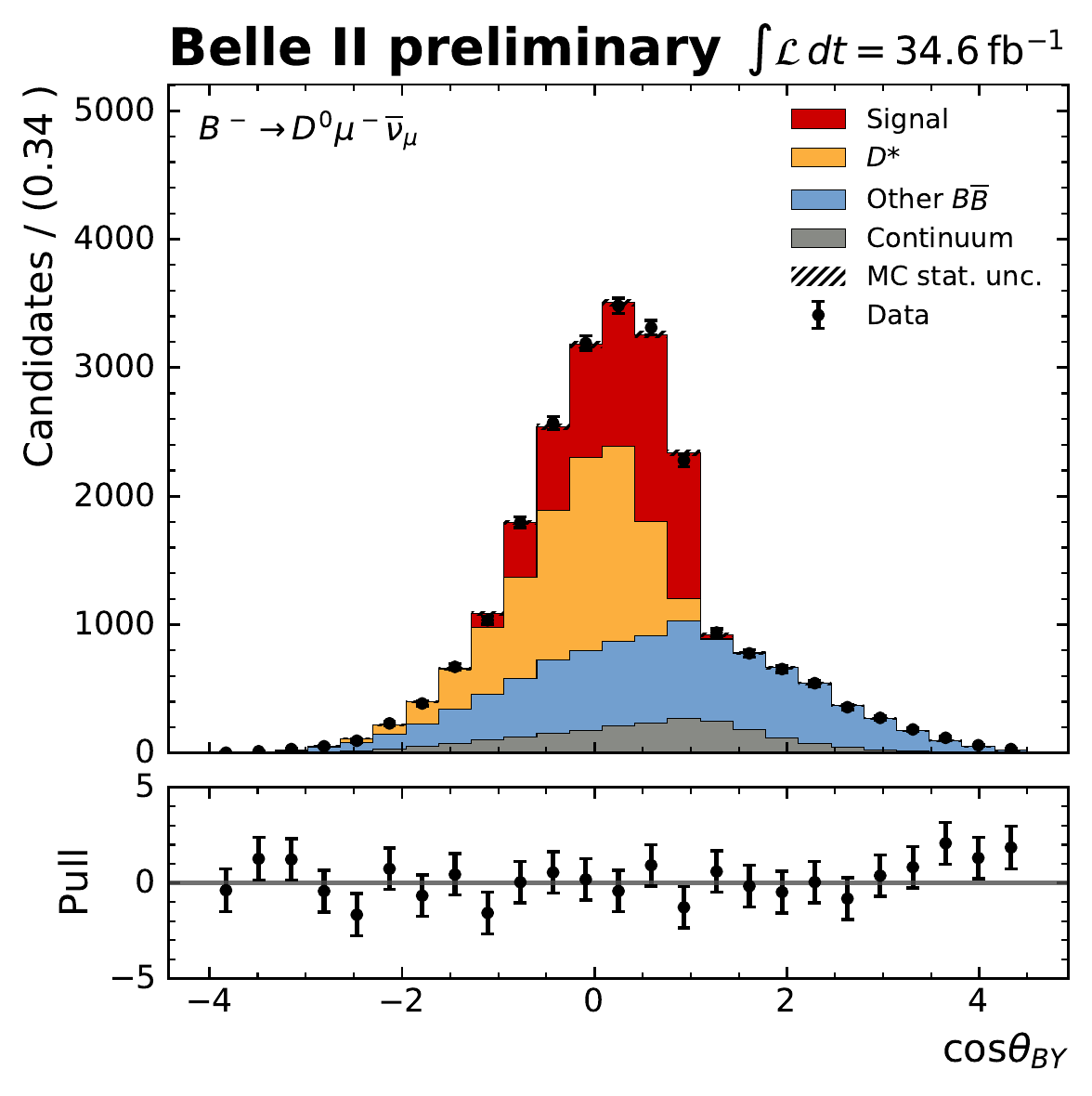}  \\
\includegraphics[width=0.5\columnwidth]{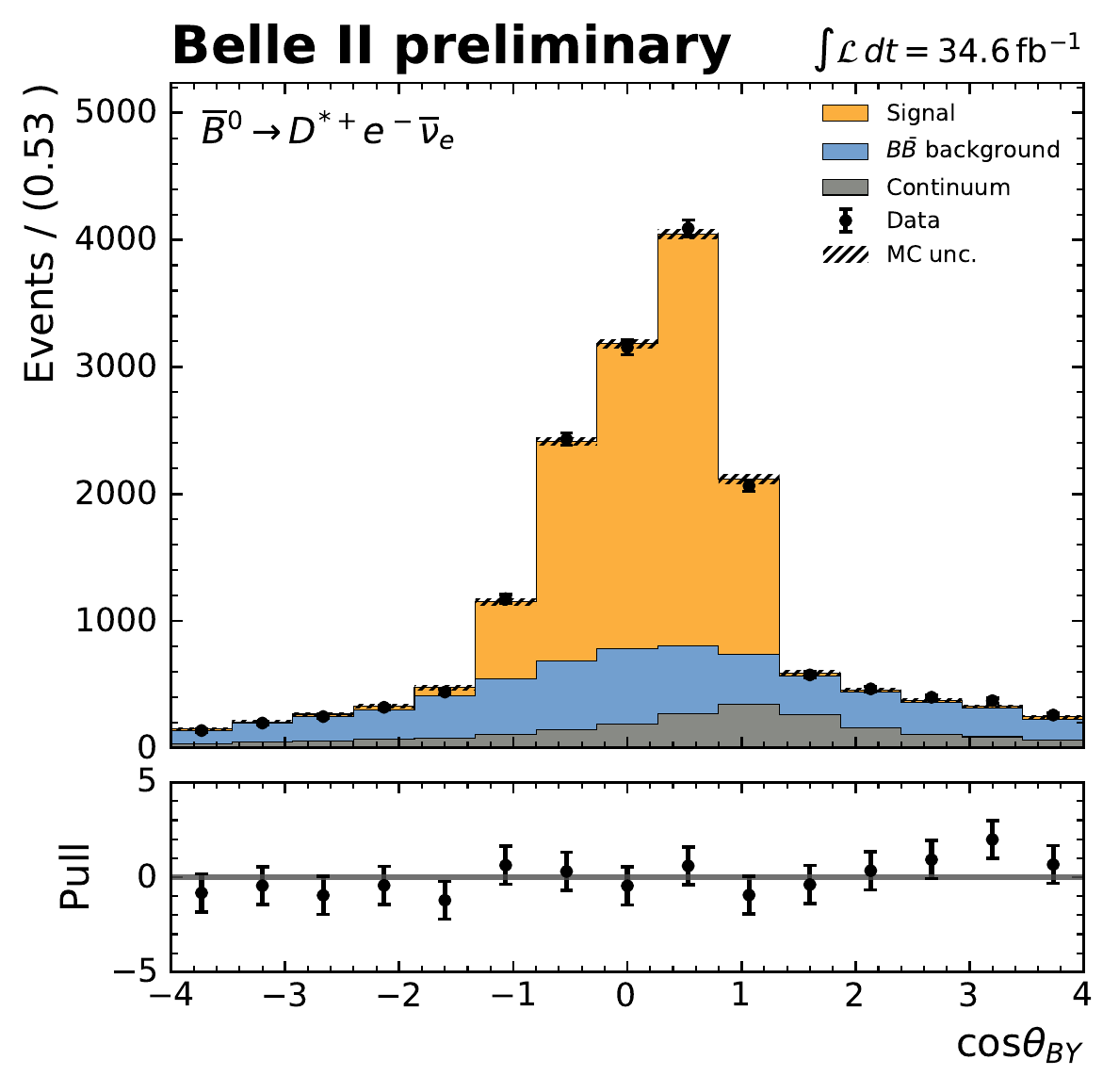} &
\includegraphics[width=0.5\columnwidth]{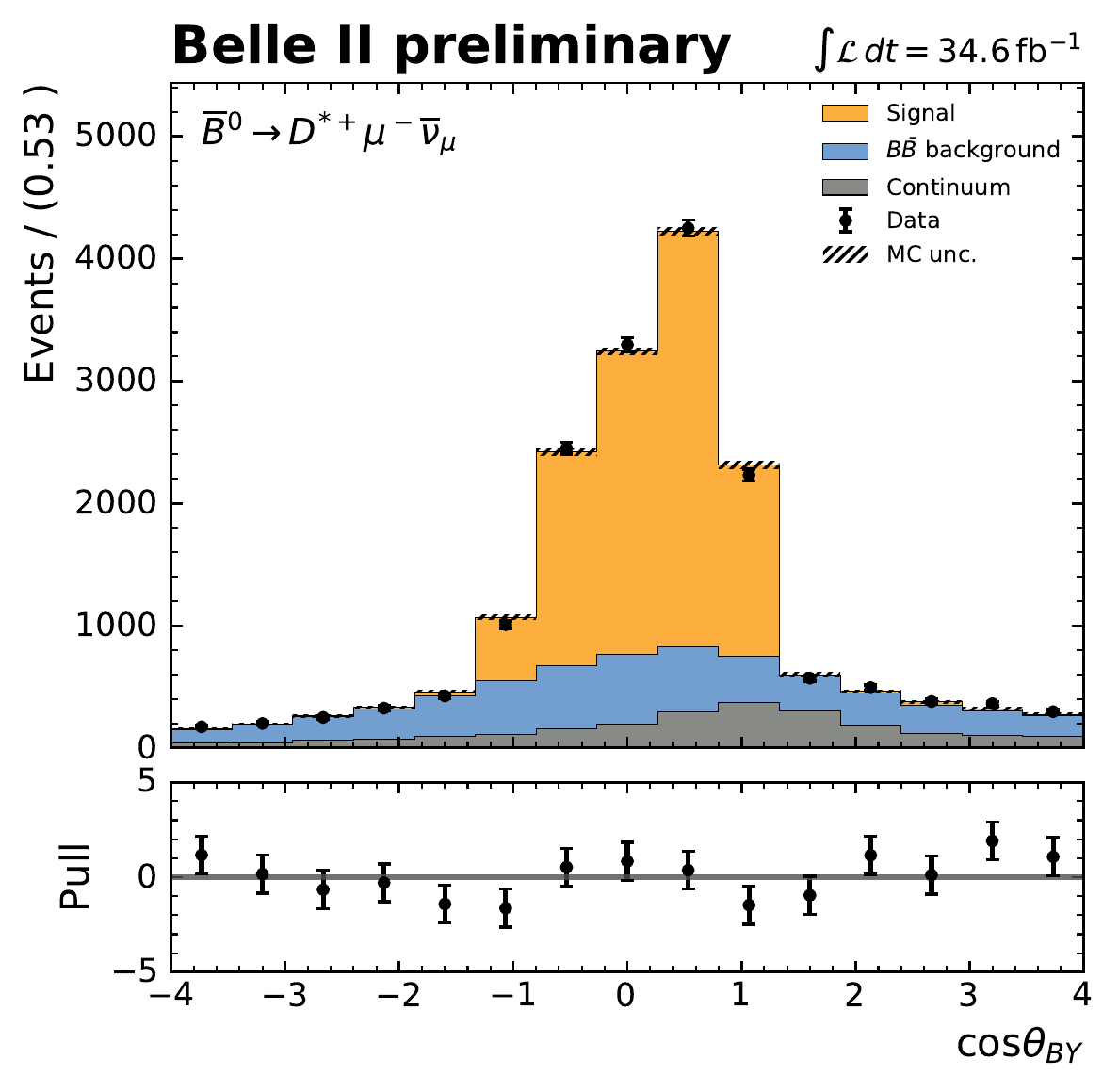}  
\end{tabular}
  \caption{The fitted \cosby\ distributions for the selected electron (left) and muon (right) candidates are shown. The top row displaying \bdlnu and the bottom row shows the results for \bdslnu. 
 }
  \label{fig:post_fit_results}
\end{center}
\end{figure}

\subsection{Branching Fraction determination for \bdslnu}

For \bdslnu we determine the measured branching fraction of the measured signal yields $N_{\mathrm{s}}$ using
\begin{align}
\BR(\bdslnu)= {N_{\mathrm{s}} \over {\epsilon \times N_{B^0}} \times \BR(D^{*+}\to D^0\pi^+)\times  \BR(D^0\to K^-\pi^+)} \, ,
\label{eq:BR-eq}
\end{align}
where $\epsilon$ is the product of the signal reconstruction efficiency and acceptance, and $N_{B^0}$ is the number of $B^0$ mesons in the data sample, further discussed in Section~\ref{sec:syst}. We determine 
\begin{align}
\resBFel \, , \label{eq:BFel}\\
\resBFmu \, . \label{eq:BFmu}
\end{align}
Both branching fractions are below, but compatible with, the current world average of $\mathcal{B}(\bdslnu) = \left( 5.05 \pm 0.14 \right) \%$ from Ref.~\cite{pdg:2020} within 0.9 and 0.8 standard deviations, respectively. The first uncertainty is from statistics and the second from systematic uncertainties, further discussed in Section~\ref{sec:syst}. The combined branching fraction is
\begin{align} 
\resBF \, ,
\end{align}
where we single out the dominant uncertainty from the slow pion efficiency. The combined branching fraction is obtained by a variance-weighted average of Eqs.~\ref{eq:BFel} and \ref{eq:BFmu}, taking into account the systematic correlations. The ratio of the electron and muon branching fraction is sensitive to lepton-flavor violating processes predicted in theories extending the Standard Model~\cite{Jung:2018lfu}. We find for the ratio
\begin{align}
    R_{e\mu} = \frac{\mathcal{B}(\benu)}{\mathcal{B}(\bmunu)} = 0.99 \pm 0.03 \, ,
\end{align}
which is compatible with the Standard Model expectation of near unity. 

\subsection{Reconstruction of the hadronic recoil parameter $w$ for \bdslnu}

For \bdslnu we reconstruct the hadronic recoil parameter $w$, defined as
\begin{align}
w & = \frac{m_B^2 + m_{D^{*+} }^2 - q^2}{ 2 m_B m_{D^{*+} }}  = v_B \cdot v_{D^{*+} } \, .
\end{align}
Here $q^2 = \left( p_B - p_{D^{*+} } \right)^2$ denotes the four-momentum transfer square of the $B$- to the $D^{*+}$-meson system. Further, $v_B$ and $v_{D^{*+} }$ denote the four-velocities of the $B$- and $D^{*+}$-mesons, respectively. Measurements of the partial branching fraction in bins of $w$ are sensitive to the non-perturbative dynamics of the \bdslnu decay and a key step to determine $\left| V_{cb} \right|$ from \bdslnu and \bdlnu decays. 

In order to reconstruct $w$, the true direction of the signal $B$ meson needs to be estimated. This is done by exploiting that the magnitude of the $B$ meson momentum vector in the CM is determined by the beam energy and its known mass. The momentum direction of the $B$ meson is constrained to lie on a cone around the momentum direction of the combined $D^{*+}\ell$ system. We combine the diamond frame reconstruction detailed in Ref.~\cite{Bevan:2014iga} with the estimated direction of the $B$ meson, as constrained by the remaining tracks and neutral clusters not used in the $D^{*+}\ell$ reconstruction (called the rest of event or ROE). This is done by modifying the diamond frame weights: cone directions opposite the ROE retain a higher weight, whereas cone directions more parallel to the ROE are weighted lower. This is implemented using weights $\frac{1}{2} \left( 1 - \widehat p_{\mathrm{ROE}} \cdot \widehat p_{\mathrm{cone}} \right)$, with $\widehat p$ denoting the normalized momentum vectors of the ROE or a cone direction. We reconstruct five bins of $w$ with bin widths larger than the expected resolution of about $0.02$. A comparison of the reconstruction resolution, comparing the reconstruction performance using the diamond frame, the estimated direction from the rest-of-the event (ROE), or the used combined approach, is shown in Fig.~\ref{fig:res}. We choose four bins with equal bin widths of $0.1$ between 1 and $1.4$, and one bin ranging from $1.4$ to $w_{\text{max}} = (m_B^2 + m_{D^{*+}}^2)/(2 m_B \, m_{D^{*+}}) = 1.504$. In each reconstructed $w$ bin, we determine the number of signal events by fitting \cosby. The post-fit distribution of the measured $w$ spectra for the electron and muon final states are shown in Fig.~\ref{fig:post_fit_w}. In Figs.~\ref{fig:fit_w_el} and \ref{fig:fit_w_mu}, the fitted \cosby\ distribution of each bin are shown.

\begin{figure}
\begin{center}
\begin{tabular}{cccc}
\includegraphics[width=0.5\columnwidth]{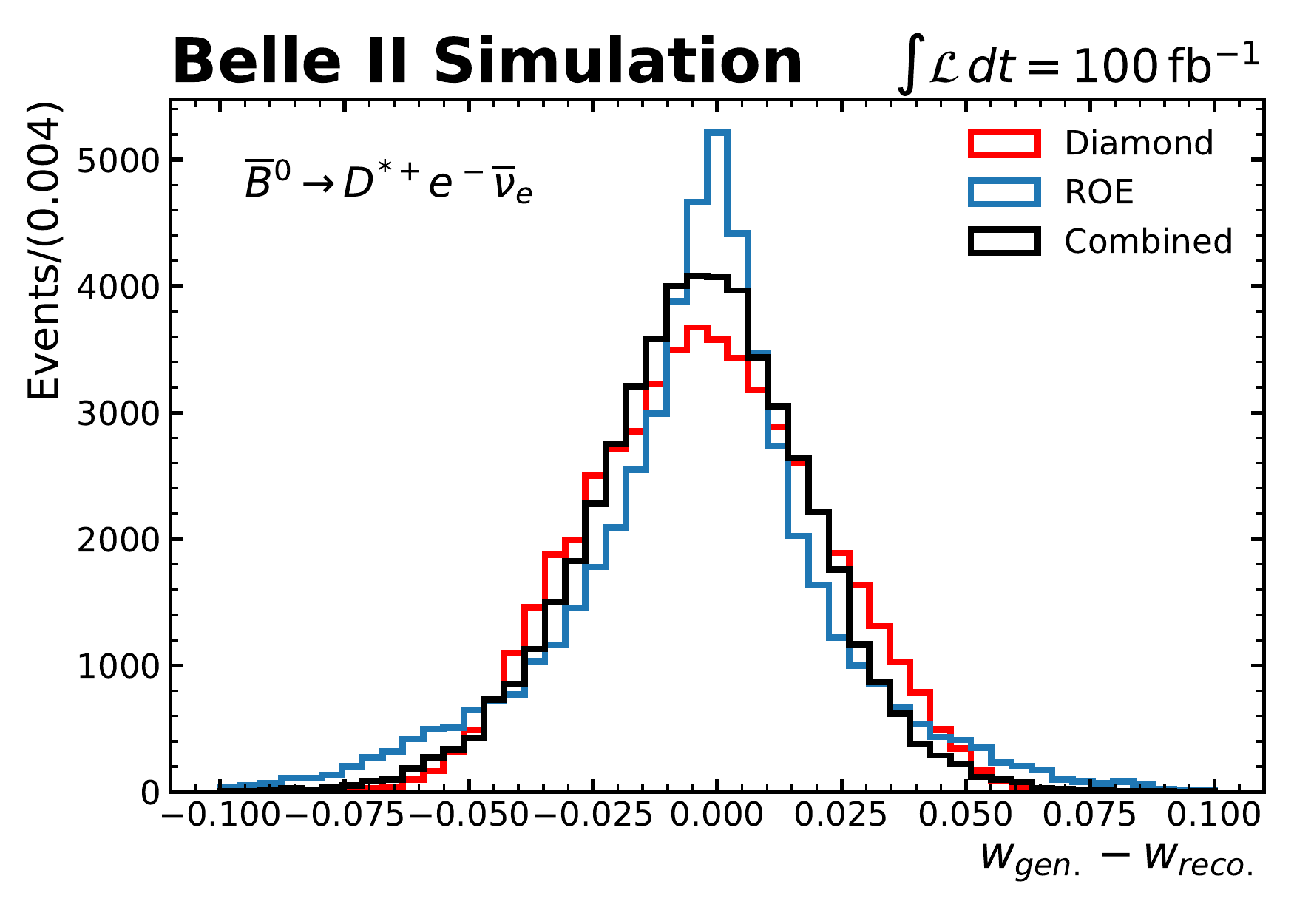} &
\includegraphics[width=0.5\columnwidth]{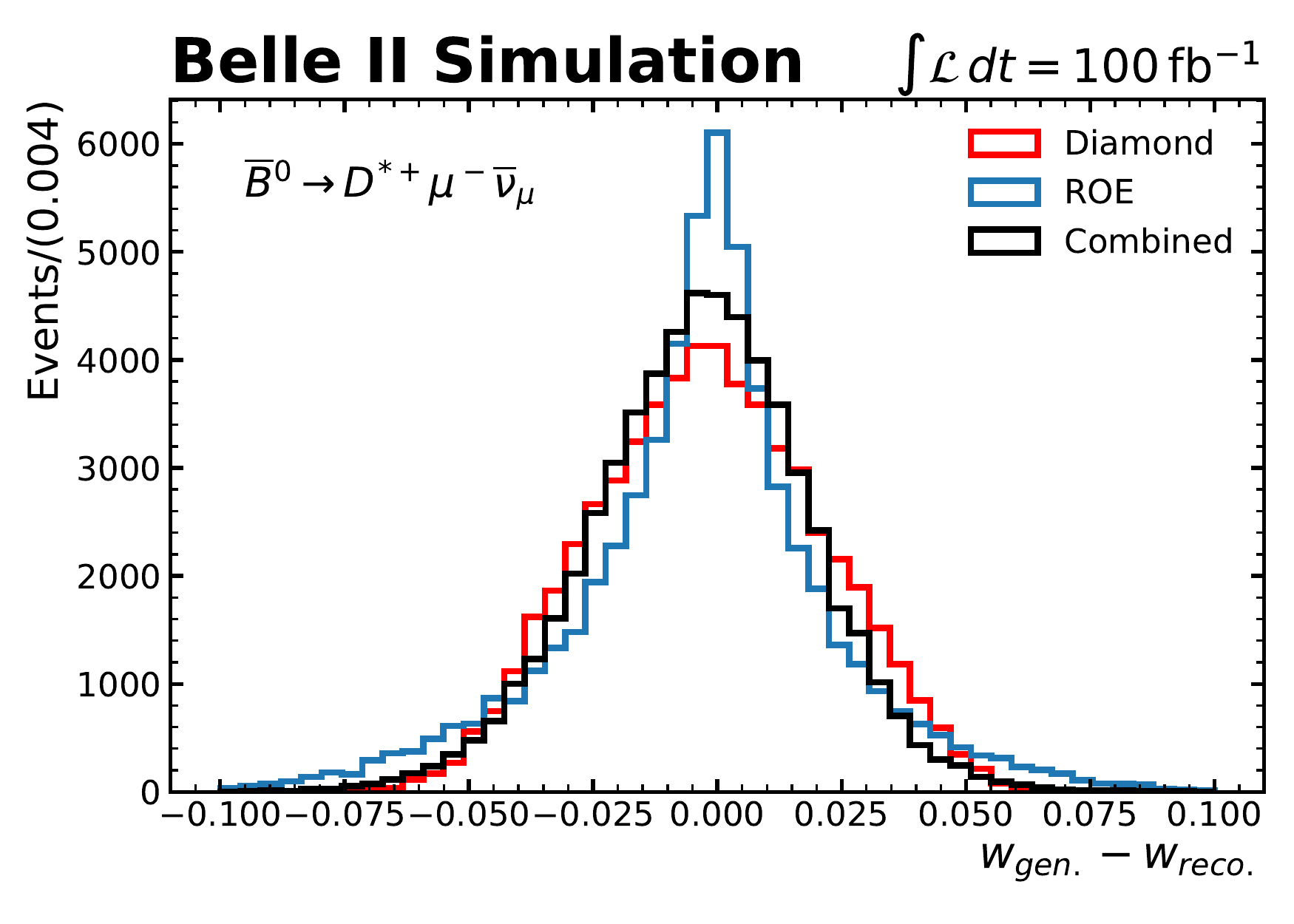} 
\end{tabular}
  \caption{ The residual of the generated and reconstructed $w$ values, after the final reconstruction and for the electron (left) and muon (right) channel, are shown. The three compared methods are: diamond frame (red), ROE (blue), and the used combined approach. For more details, see text. 
 }
  \label{fig:res}
\end{center}
\end{figure}

\begin{figure}
\begin{center}
\begin{tabular}{cccc}
\includegraphics[width=0.5\columnwidth]{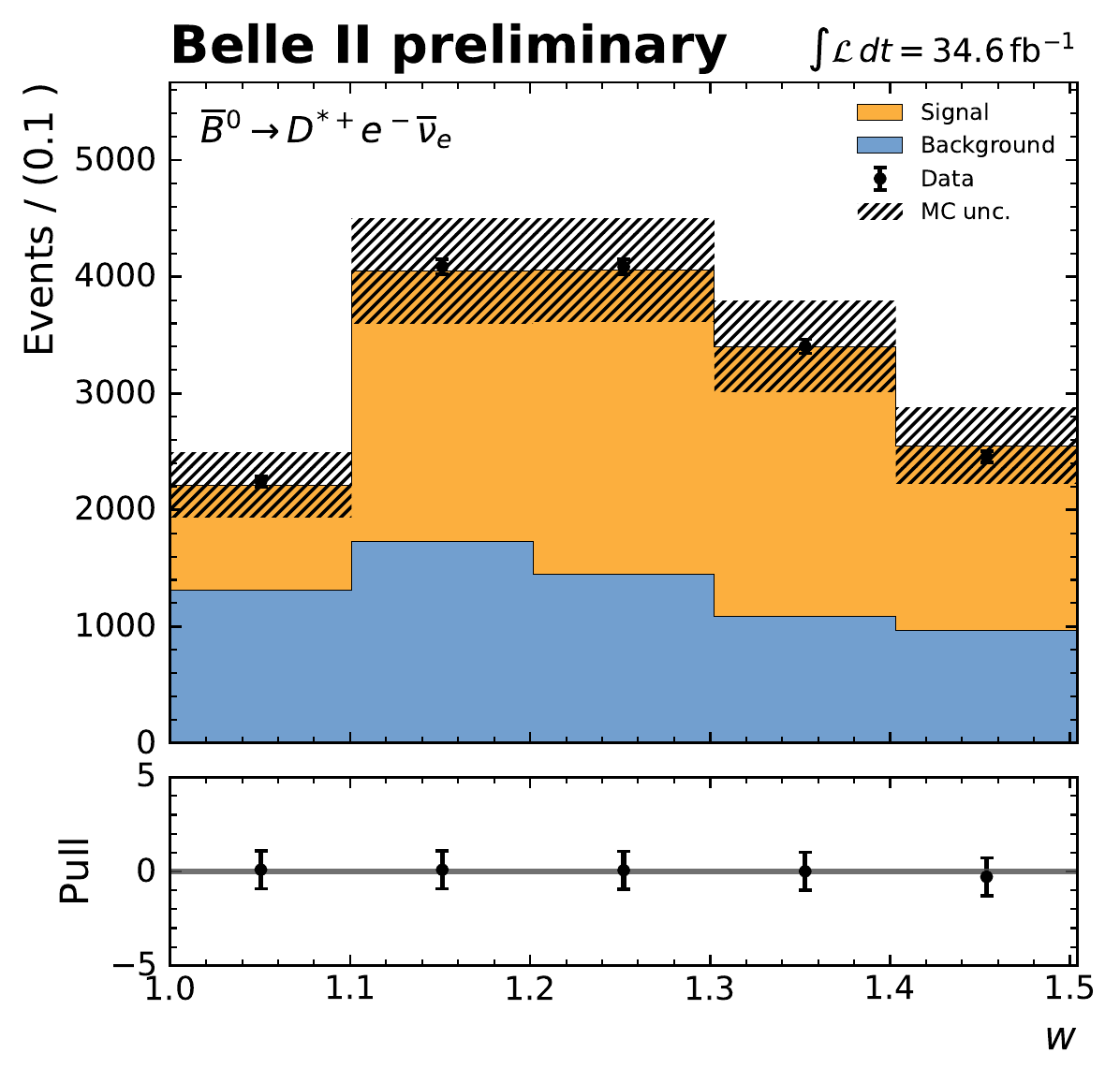} &
\includegraphics[width=0.5\columnwidth]{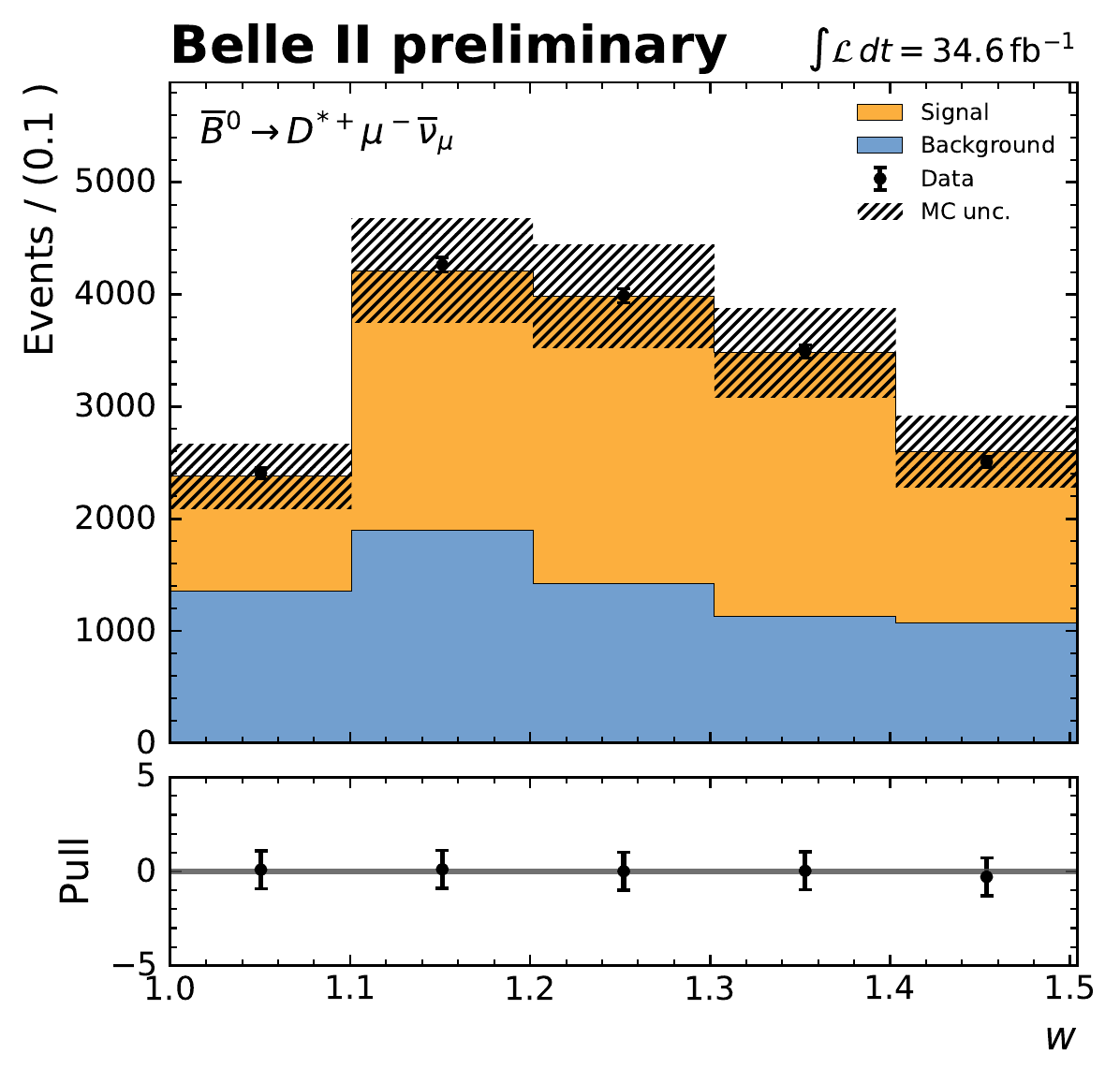} 
\end{tabular}
  \caption{The fitted $w$ distribution for electron (left) and muon (right) \bdslnu candidates are shown, after fitting \cosby\ in each bin. The background can be described adequately as can be seen by the near zero pulls in each bin. 
 }
  \label{fig:post_fit_w}
\end{center}
\end{figure}

\begin{figure}
\begin{center}
\includegraphics[width=0.32\columnwidth]{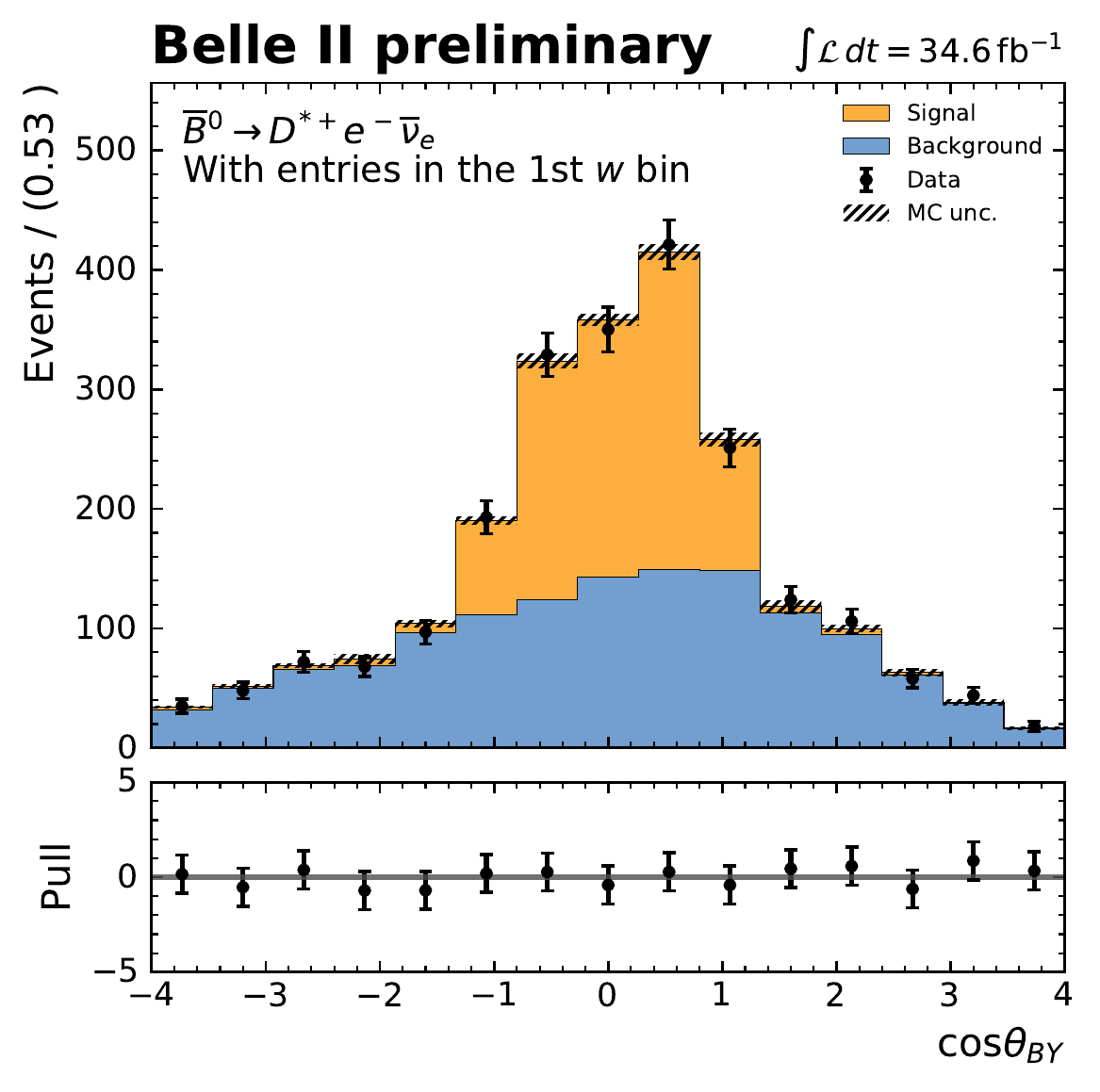} 
\includegraphics[width=0.32\columnwidth]{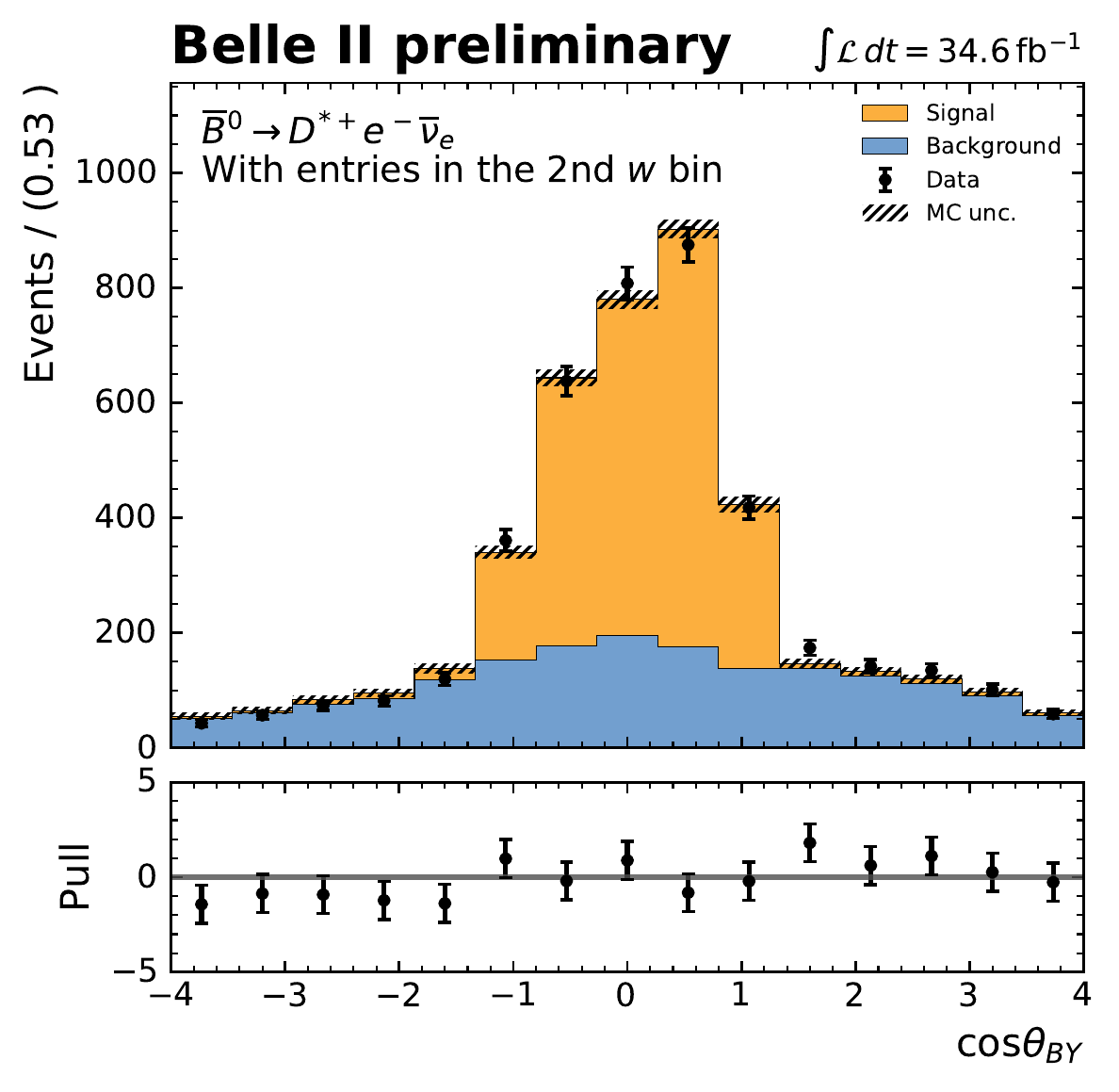}  
\includegraphics[width=0.32\columnwidth]{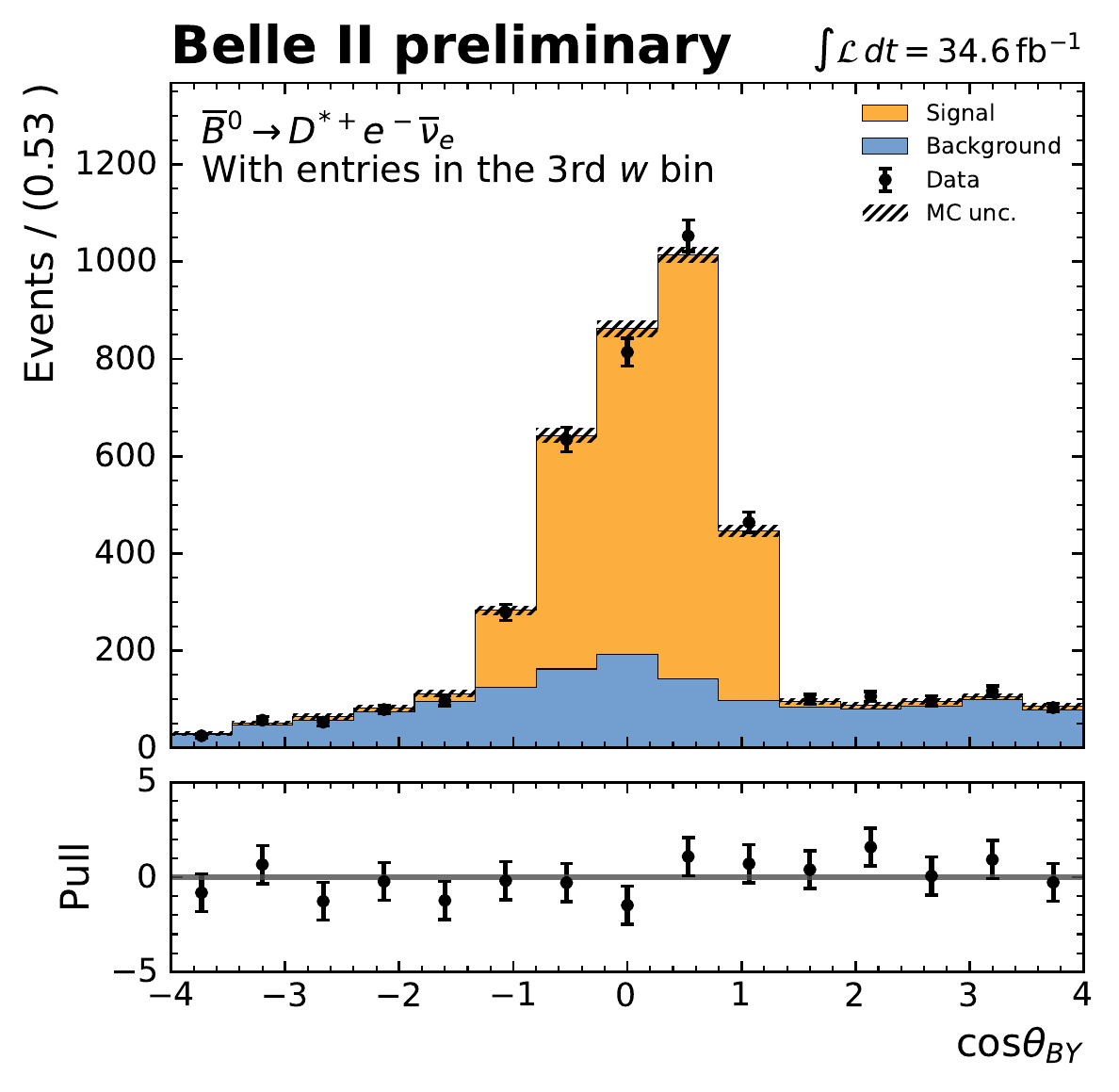}  \\
\includegraphics[width=0.32\columnwidth]{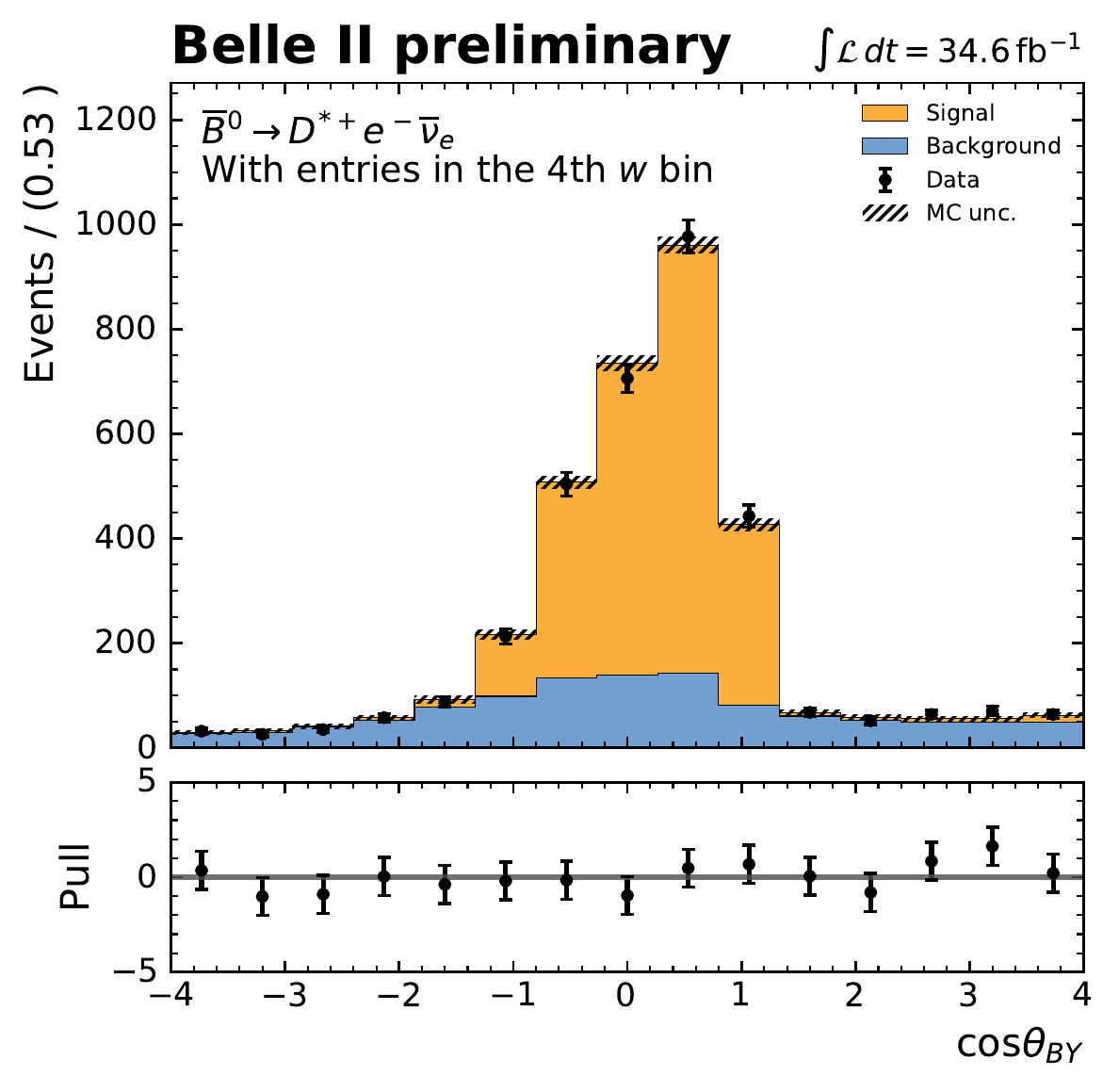}  
\includegraphics[width=0.32\columnwidth]{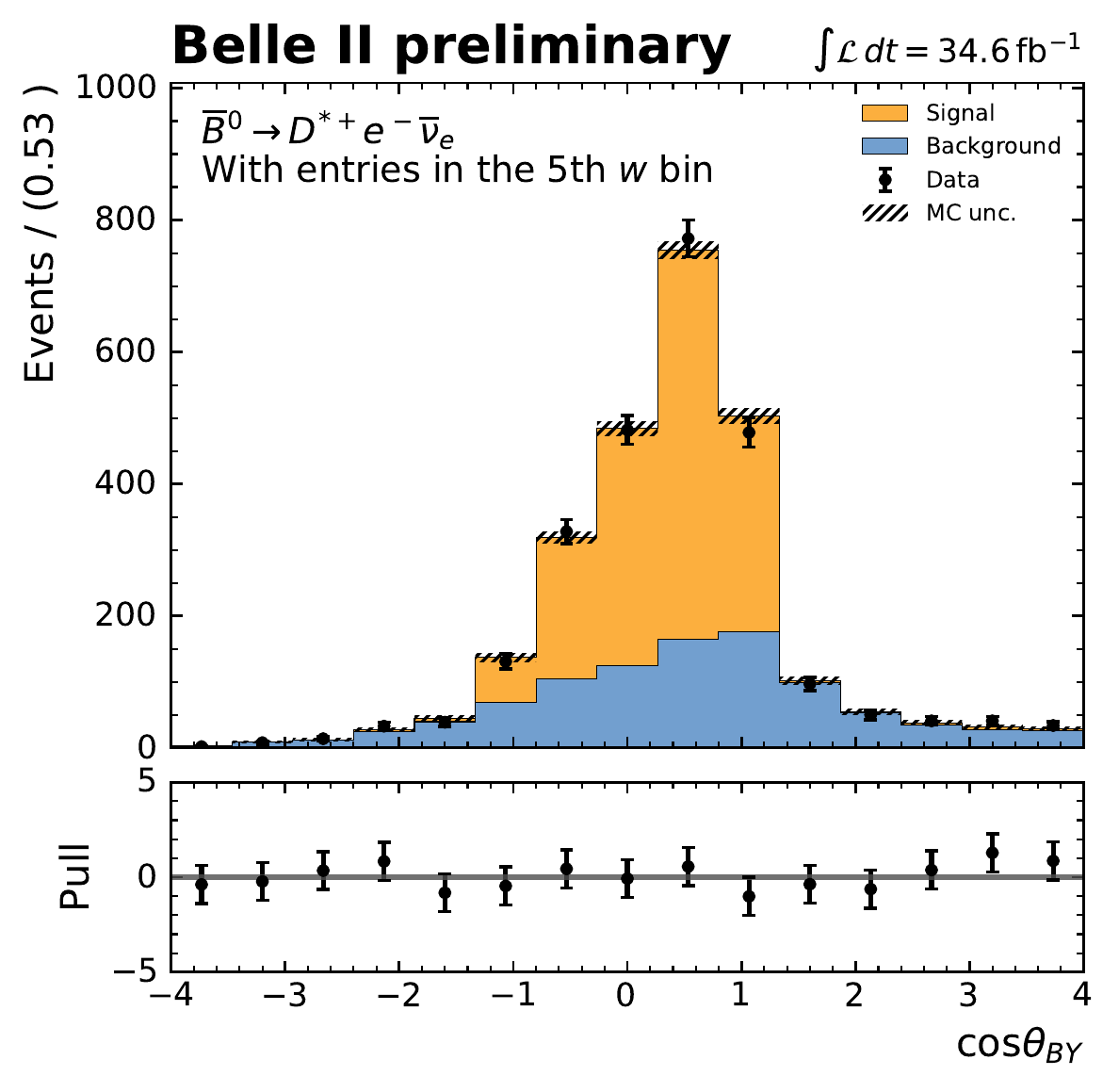} 
  \caption{The fitted \cosby\ distributions of all $w$ bins of \benu for the electron final state are shown. 
 }
  \label{fig:fit_w_el}
\end{center}
\end{figure}

\begin{figure}
\begin{center}
\includegraphics[width=0.32\columnwidth]{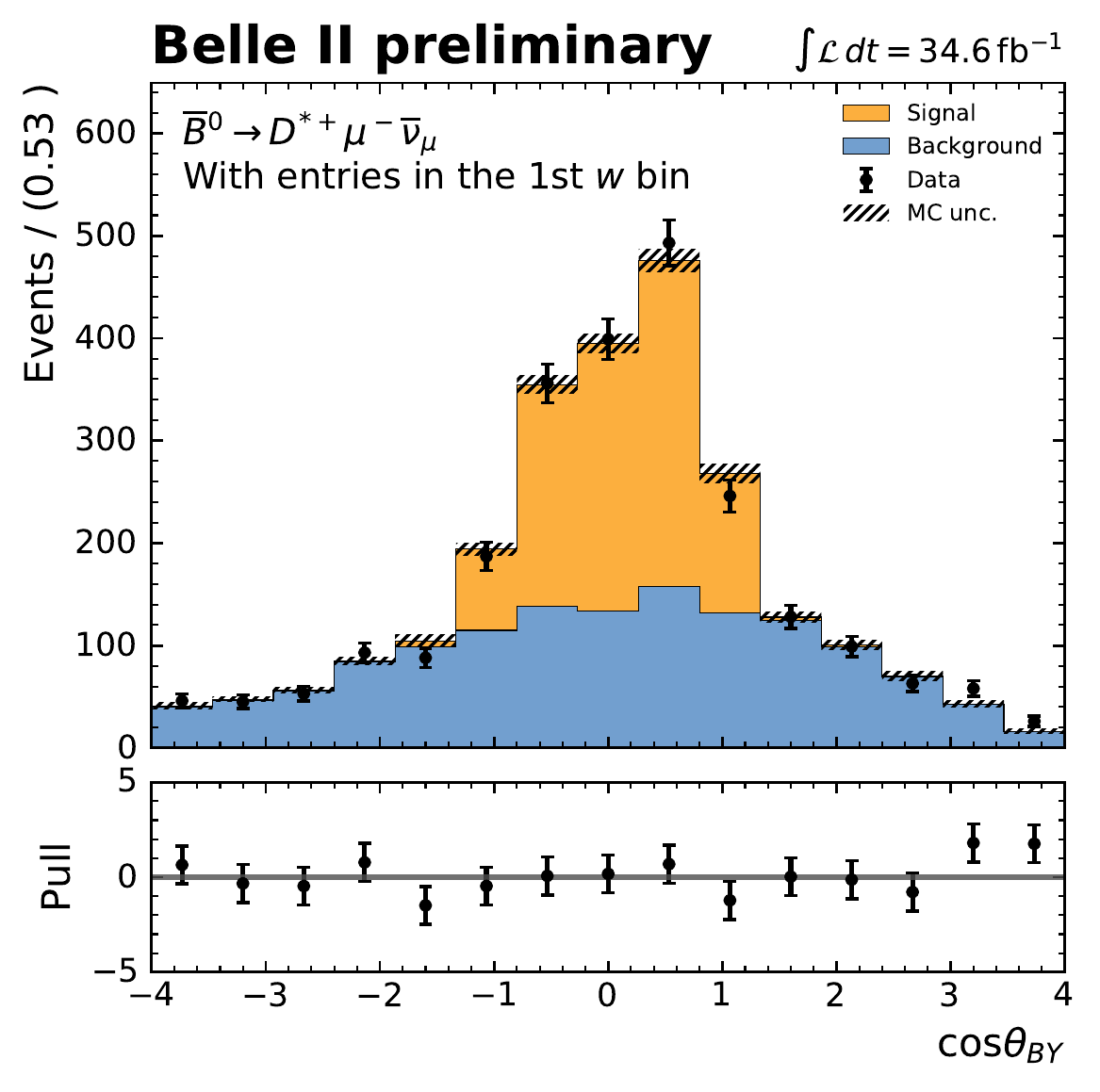} 
\includegraphics[width=0.32\columnwidth]{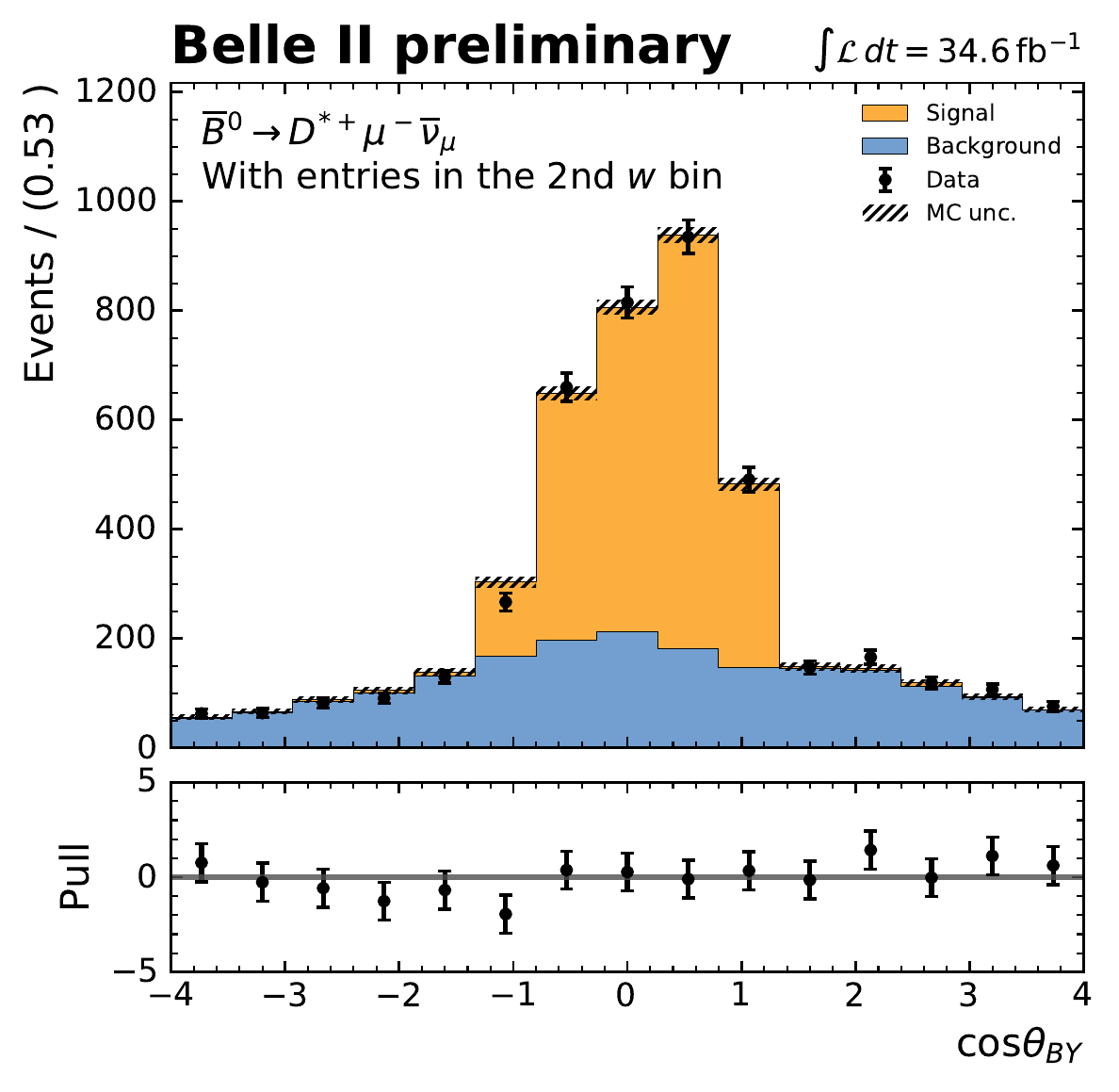}  
\includegraphics[width=0.32\columnwidth]{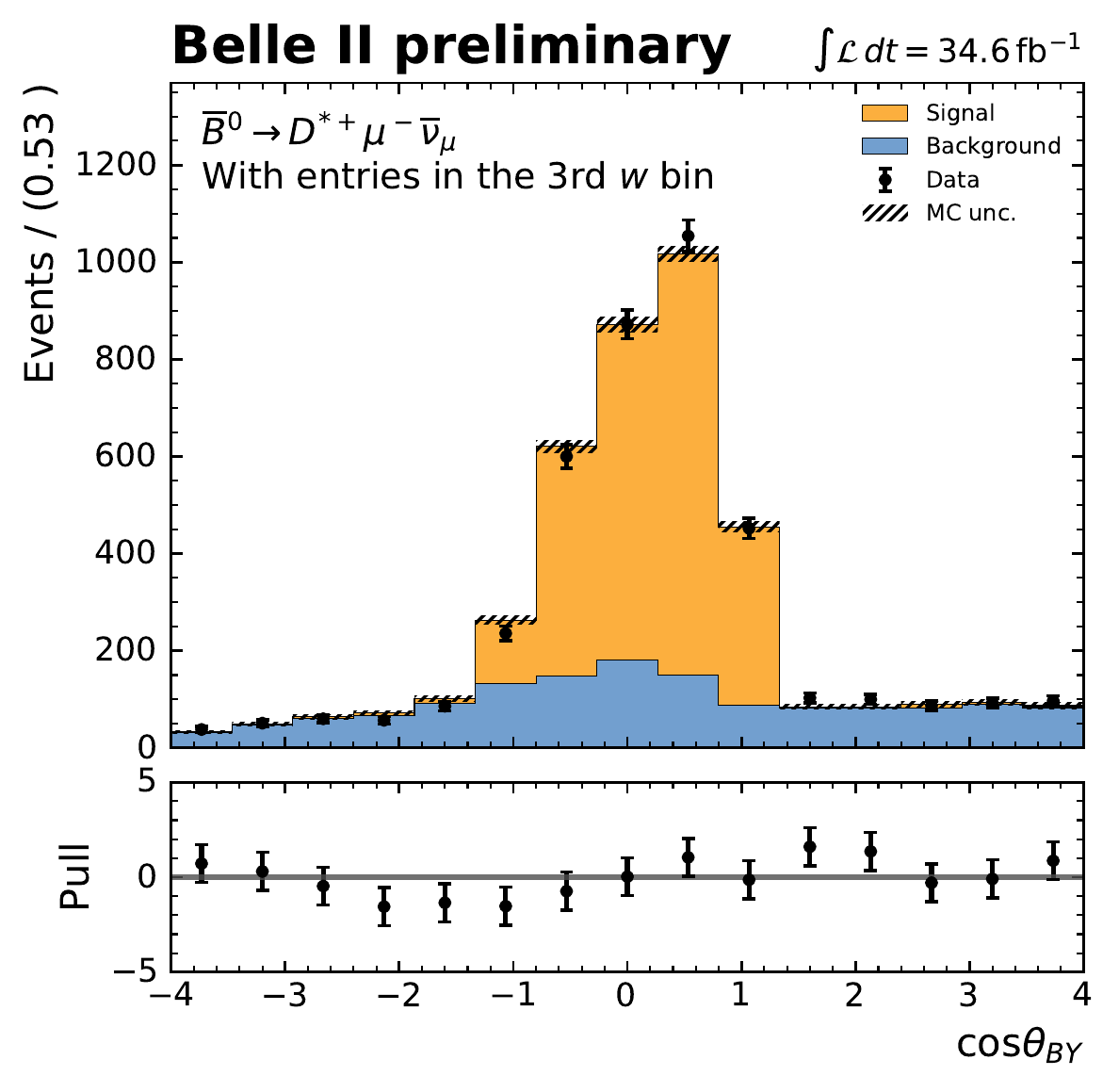}  \\
\includegraphics[width=0.32\columnwidth]{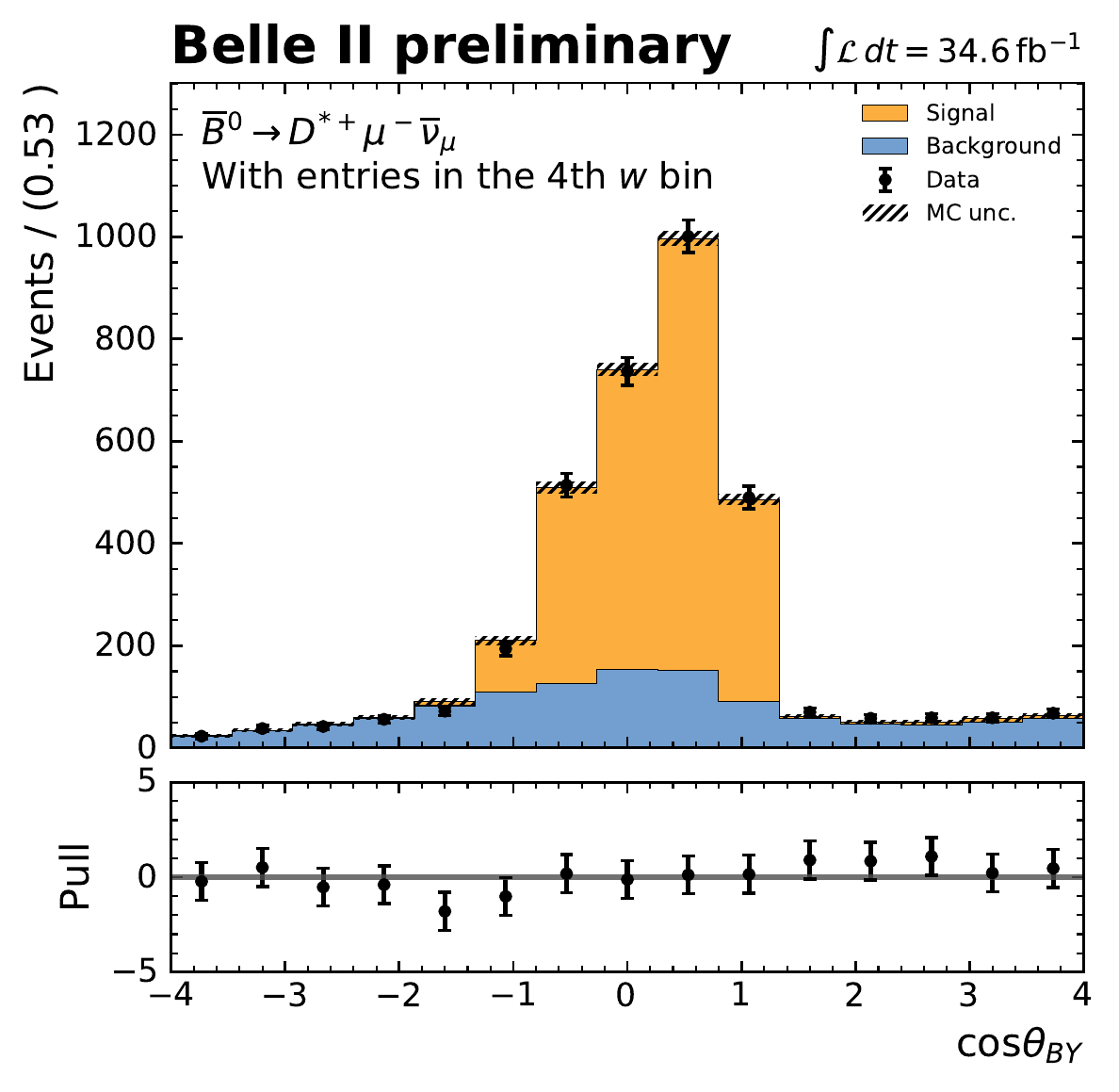}  
\includegraphics[width=0.32\columnwidth]{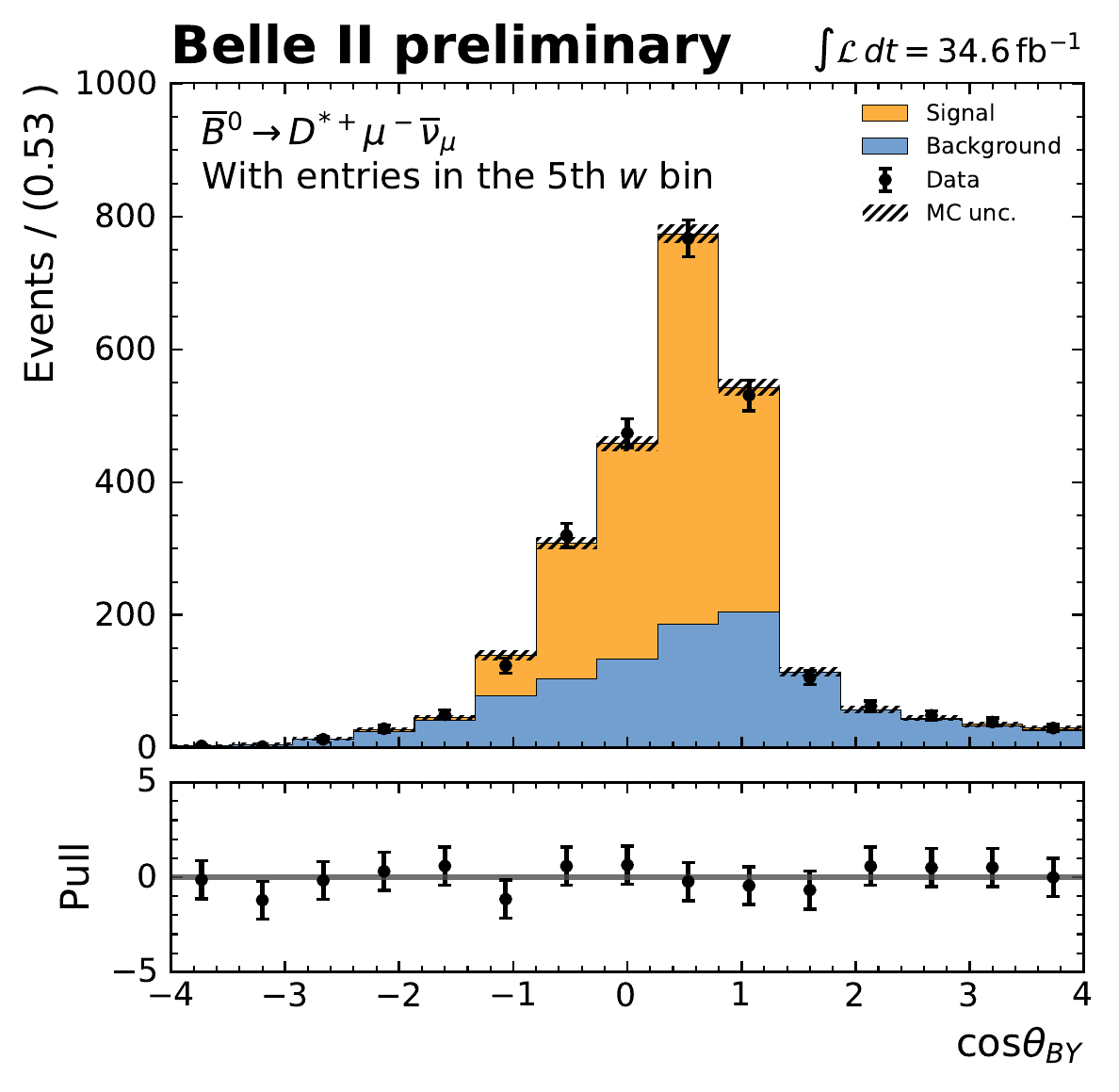} 
  \caption{The fitted \cosby\ distributions of all $w$ bins of \bmunu for the muon final state are shown. 
 }
  \label{fig:fit_w_mu}
\end{center}
\end{figure}

\subsection{Unfolding of the hadronic recoil parameter $w$ for \bdslnu}

In order to confront the measured $w$ distributions with predictions for the decay rate, effects from resolution and efficiencies have to be reverted. This is done by constructing a $\chi^2$ function of the form
\begin{align}
 \chi^2 = \left( \bold{ N_{\mathrm{s}} } - \bold{ \overline N_{\mathrm{s}} } \times  \mathcal{M} \right) C^{-1}_{\mathrm{exp}}  \left( \bold{ N_{\mathrm{s}}} - \bold{ \overline N_{\mathrm{s}}} \times  \mathcal{M} \right) \, .
\end{align}
Here, $C_{\mathrm{exp}}$ denotes the experimental covariance of the measurement. The migration matrix $\mathcal{M}$ denotes the conditional probabilities 
\begin{align}
 \mathcal{M}_{ij} & = \mathcal{P}( \text{measured value in bin $i$} | \text{true value in bin $j$} ) \, ,
\end{align}
mapping the reconstructed signal yields $\bold{ N_s}$, expressed as a vector of the bins, into their unfolded values $ \bold{ \overline N_s}$. The unfolded yields are converted into partial decay rates using 
\begin{align}
\Delta \Gamma_i= { \overline N_{\mathrm{s}\, i} \times \tau_{B^0} \over {\epsilon_i \times N_{B^0}} \times \BR(D^{*+}\to D^0\pi^+)\times  \BR(D^0\to K^-\pi^+)} \, ,
\label{eq:dG}
\end{align}
with $\tau_{B^0} = \left(1.519 \pm 0.004  \right)$ ps the $B^0$ meson lifetime. Further, $\epsilon_i$ denotes the reconstruction efficiency and acceptance of signal events with true values of $w$ in bin $i$. The resulting unfolded distributions are shown in Fig.~\ref{fig:bgl_v_meas} and compared to the BGL form factor parameters of Refs.~\cite{Grinstein:2017nlq,Bigi:2017njr}.

\begin{figure}
\begin{center}
\includegraphics[width=0.45\columnwidth]{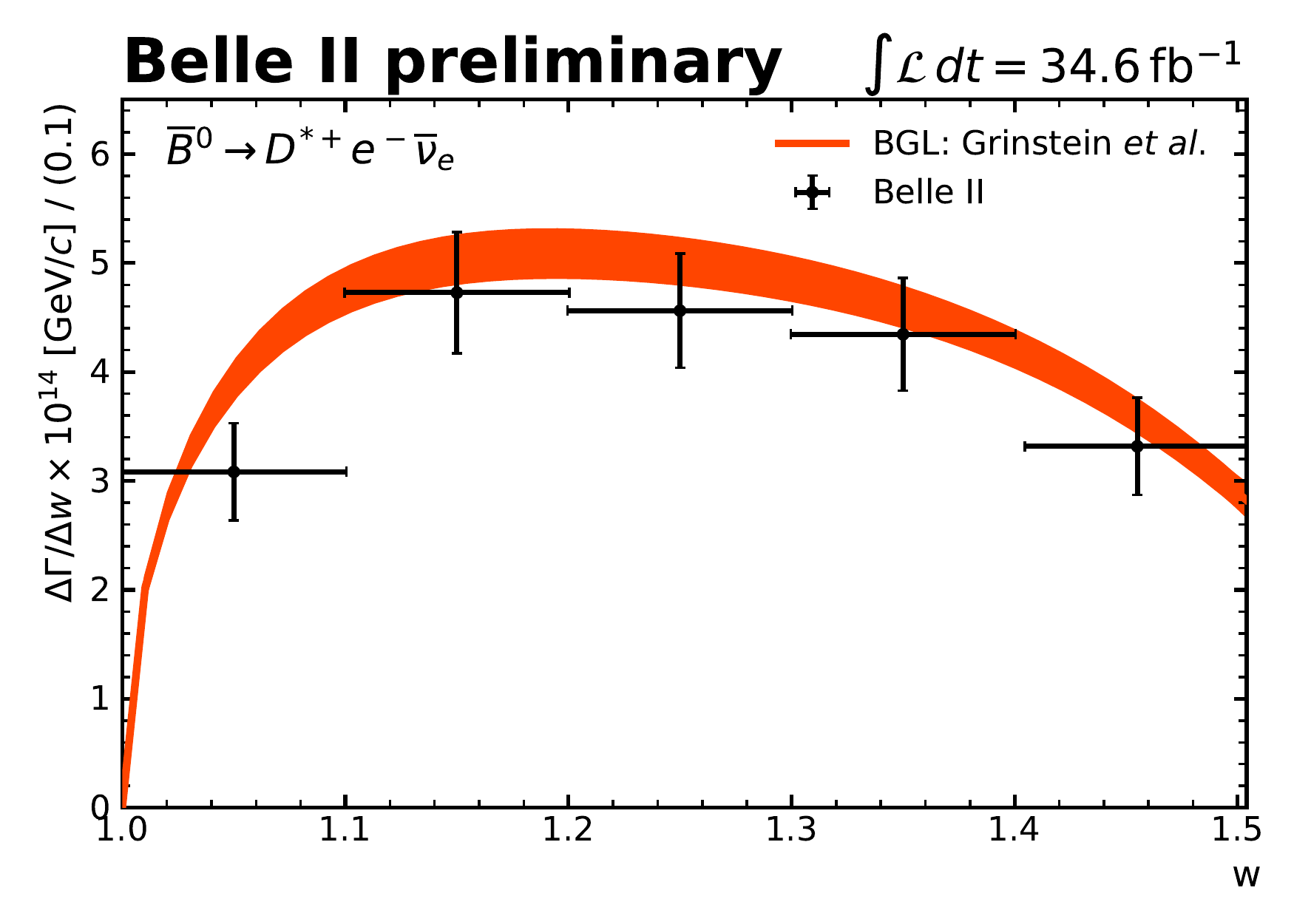} 
\includegraphics[width=0.45\columnwidth]{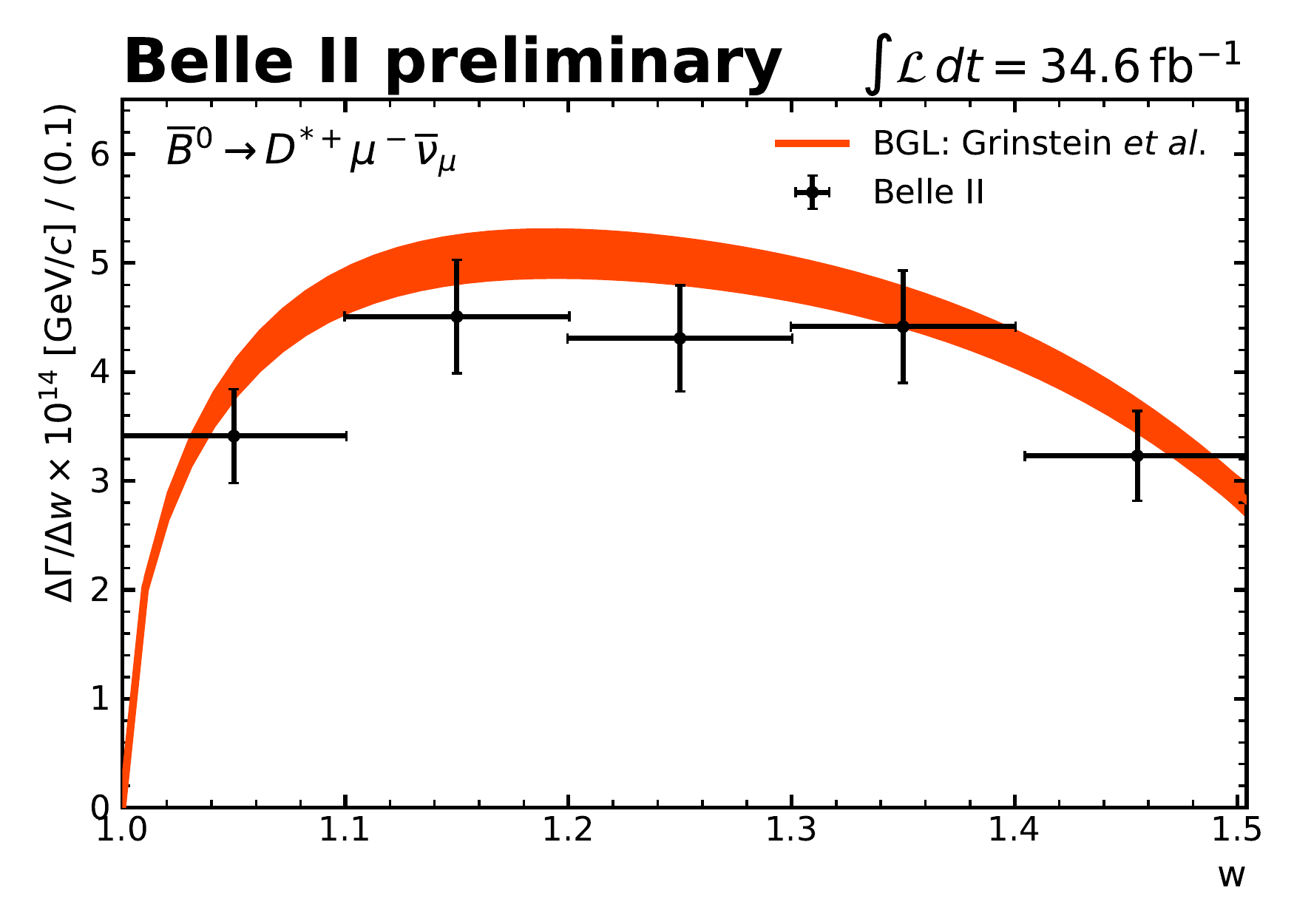} 
  \caption{The measured partial decay rates for electrons and muons are compared to the BGL form factor parameters of Refs.~\cite{Bigi:2017njr,Grinstein:2017nlq}.
 }
  \label{fig:bgl_v_meas}
\end{center}
\end{figure}

%%%%%%%%%%%%%%%%%%%%%%%%%
\section {Systematic uncertainties}\label{sec:syst}

The relative systematic uncertainties affecting the \bdslnu branching fraction measurement are listed in Table~\ref{tab:syst}. We assume no correlation among the individual sources of uncertainty and sum them in quadrature to obtain the total systematic uncertainty. The  methods used for obtaining these uncertainties are detailed below.

\begin{table}[htbp]
\begin{center}
\begin{tabular}{l|c|c}\hline\hline
Source & \multicolumn{2}{c}{Relative uncertainty (\%)} \\
    \cline{2-3}
       &\benu & \bmunu \\
\hline
PDF shape uncertainties  & 0.7 & 0.6 \\
$\BR(\bar B\to D^{**} \ell\bar\nu)$ & 0.1  & $<$ 0.1 \\
Lepton-ID & 0.4 & 1.9 \\
MC statistics, efficiency  & $<$ 0.1  & $<$ 0.1 \\
Tracking of $K$, $\pi$, $\ell$ & 2.4 & 2.4 \\
Tracking of $\pis$ & 9.9 & 9.9 \\
$N_{B^0}$  & 2.0 & 2.0
\\
Charm branching fractions & 1.1  & 1.1 \\
 \bdslnu Form Factors & 1.1  & 1.1 \\
\hline
Total &10.5 & 10.7\\
 \hline\hline
\end{tabular}
\end{center}
\caption{Summary of the relative systematic uncertainties for the measurements of $\BR(\bdslnu)$.
The first two uncertainties impact the extracted signal yield, while the others impact the other factors of Eq.~(\ref{eq:BR-eq}).
}\label{tab:syst}
\end{table}

The lepton-identification corrections are measured with statistical uncertainties that arise from the limited size of the control samples, as well as systematic uncertainties. We produce 500 sets of correction values sampled from Gaussian distributions that reflect these uncertainties, accounting for systematic correlations. Each set of corrections is used to estimate the uncertainty on the efficiencies and on the \cosby\ distributions.

The impact of the finite sizes of the MC samples is directly incorporated into the fit procedure via nuisance parameters. 

The semileptonic decays $\bar B\to D^{**} \ell\bar\nu$, where $D^{**}$ indicates an excited charm meson heavier than the $D^*$, have a similar particle content to that of signal decays. As a result, the fit may be biased if the branching fractions of $\bar B\to D^{**} \ell\bar\nu$ are incorrect in the generic MC sample. To estimate the systematic uncertainty, we obtain the $B\bar B$ PDF from the MC after varying the branching fractions for these decays by $\pm 25\%$, which is twice the relative uncertainty on $\BR(\bar B\to D^0 \pi^+ \ell^-\bar\nu)$. The resulting change in the signal yield is taken as the systematic uncertainty.

The tracking efficiency uncertainty for the lepton, kaon, and pion is 0.80\% per track. This is obtained by comparing $R_{2/3}$ for $e^+e^-\to \tau^+\tau^-$ events in data and MC, where $R_{2/3}$ is the fraction of 3-prong $\tau$ decays in which only two hadron tracks are found. The uncertainty on the soft pion tracking efficiency is determined by the study of $B \to D^* \, \pi$ and  $B \to D^* \, \rho$ decays and estimated to be 9.9\%. 

To obtain the number of $B^0$ mesons in the sample, we use the relation 

\begin{equation}
N_{B^0} =  2 \times N_{B \bar B} \times \left( 1 + f_{+0} \right)^{-1} \, . 
\end{equation}
Here $f^{+0} = \mathcal{B}(\Upsilon(4S) \to B^+ \, B^-) /  \mathcal{B}(\Upsilon(4S) \to B^0 \, \overline B^0) = 1.058 \pm 0.024$~\cite{pdg:2020}. The number of $B$ meson pairs in the analyzed data set is determined to be \NBB. 

The uncertainties of the selection efficiencies on the used form factors used to simulate \bdslnu are taken from Refs.~\cite{Bigi:2017njr,Grinstein:2017nlq} and varied within their uncertainties. 

Lastly, we account for the impact of the uncertainties in the charm branching fractions, $\BR(\Dstarp \to \Dz \pip)$ = $(67.7\pm0.5)\%$ and $\BR(\Dz \to \Km \pip)$ = $(3.950\pm0.031)\%$~\cite{Zyla:2020}, on the signal branching fraction.

%%%%%%%%%%%%%%
\section{Summary and Conclusions}

We present measurements of the semileptonic \bdslnu and \bdlnu processes using \lumi of recorded collision events of Belle~II data. We demonstrate the capability to reconstruct and separate \bdlnu candidates from the large backgrounds from \bdslnu and other processes. In addition, we measure the \bdslnu branching fraction and obtain a value of 
\begin{align}
\resBF \, ,
\end{align}
lower than, but in good agreement with, the current world average. The largest systematic uncertainty stems from the knowledge of the slow pion reconstruction efficiency. This uncertainty will improve with the statistics of the control samples that will become available soon. In addition, we demonstrate the capability to reconstruct the hadronic recoil parameter $w$ and present unfolded partial decay rates. Such measurements in both \bdslnu and \bdlnu are crucial for future precision measurements of $|V_{cb}|$ in these channels by Belle~II.

\section{ACKNOWLEDGEMENTS}
We thank the SuperKEKB group for the excellent operation of the
accelerator; the KEK cryogenics group for the efficient
operation of the solenoid; and the KEK computer group for
on-site computing support.
This work was supported by the following funding sources:
%Armenia
Science Committee of the Republic of Armenia Grant No. 18T-1C180;
%Australia
Australian Research Council and research grant Nos.
DP180102629, 
DP170102389, 
DP170102204, 
DP150103061, 
FT130100303, 
and
FT130100018; 
%Austria
Austrian Federal Ministry of Education, Science and Research, and
Austrian Science Fund No. P 31361-N36; 
%Canada
Natural Sciences and Engineering Research Council of Canada, Compute Canada and CANARIE;
%China
Chinese Academy of Sciences and research grant No. QYZDJ-SSW-SLH011,
National Natural Science Foundation of China and research grant Nos.
11521505,
11575017,
11675166,
11761141009,
11705209,
and
11975076,
LiaoNing Revitalization Talents Program under contract No. XLYC1807135,
Shanghai Municipal Science and Technology Committee under contract No. 19ZR1403000,
Shanghai Pujiang Program under Grant No. 18PJ1401000,
and the CAS Center for Excellence in Particle Physics (CCEPP);
%Czech Republic
the Ministry of Education, Youth and Sports of the Czech Republic under Contract No.~LTT17020 and 
Charles University grants SVV 260448 and GAUK 404316;
%EU
European Research Council, 7th Framework PIEF-GA-2013-622527, 
Horizon 2020 Marie Sklodowska-Curie grant agreement No. 700525 `NIOBE,' 
and
Horizon 2020 Marie Sklodowska-Curie RISE project JENNIFER2 grant agreement No. 822070 (European grants);
%France
L'Institut National de Physique Nucl\'{e}aire et de Physique des Particules (IN2P3) du CNRS (France);
%Germany
BMBF, DFG, HGF, MPG, AvH Foundation, and Deutsche Forschungsgemeinschaft (DFG) under Germany's Excellence Strategy -- EXC2121 ``Quantum Universe''' -- 390833306 (Germany);
%India
Department of Atomic Energy and Department of Science and Technology (India);
%Israel
Israel Science Foundation grant No. 2476/17
and
United States-Israel Binational Science Foundation grant No. 2016113;
%Italy
Istituto Nazionale di Fisica Nucleare and the research grants BELLE2;
%Japan
Japan Society for the Promotion of Science,  Grant-in-Aid for Scientific Research grant Nos.
16H03968, 
16H03993, 
16H06492,
16K05323, 
17H01133, 
17H05405, 
18K03621, 
18H03710, 
18H05226,
19H00682, % Niigata
26220706,
and
26400255,
the National Institute of Informatics, and Science Information NETwork 5 (SINET5), 
and
the Ministry of Education, Culture, Sports, Science, and Technology (MEXT) of Japan;  
%Korea
National Research Foundation (NRF) of Korea Grant Nos.
2016R1\-D1A1B\-01010135,
2016R1\-D1A1B\-02012900,
2018R1\-A2B\-3003643,
2018R1\-A6A1A\-06024970,
2018R1\-D1A1B\-07047294,
2019K1\-A3A7A\-09033840,
and
2019R1\-I1A3A\-01058933,
Radiation Science Research Institute,
Foreign Large-size Research Facility Application Supporting project,
the Global Science Experimental Data Hub Center of the Korea Institute of Science and Technology Information
and
KREONET/GLORIAD;
%Malaysia
Universiti Malaya RU grant, Akademi Sains Malaysia and Ministry of Education Malaysia;
%Mexico
% CINVESTAV-IPN, UNAM, UAS, BUAP and CONACYT are funded under
Frontiers of Science Program contracts
FOINS-296,
CB-221329,
CB-236394,
CB-254409,
and
CB-180023, and SEP-CINVESTAV research grant 237 (Mexico);
%Poland
the Polish Ministry of Science and Higher Education and the National Science Center;
%Russia
the Ministry of Science and Higher Education of the Russian Federation,
Agreement 14.W03.31.0026;
%Saudi Arabia
University of Tabuk research grants
S-1440-0321, S-0256-1438, and S-0280-1439 (Saudi Arabia);
%Slovenia
Slovenian Research Agency and research grant Nos.
J1-9124
and
P1-0135; 
%Spain
Agencia Estatal de Investigacion, Spain grant Nos.
FPA2014-55613-P
and
FPA2017-84445-P,
and
CIDEGENT/2018/020 of Generalitat Valenciana;
%Taiwan
Ministry of Science and Technology and research grant Nos.
MOST106-2112-M-002-005-MY3
and
MOST107-2119-M-002-035-MY3, 
and the Ministry of Education (Taiwan);
%Thailand
Thailand Center of Excellence in Physics;
%Turkey
TUBITAK ULAKBIM (Turkey);
%Ukraine
Ministry of Education and Science of Ukraine;
%USA
the US National Science Foundation and research grant Nos.
PHY-1807007 % Luther
and
PHY-1913789, % Indiana CEEM
and the US Department of Energy and research grant Nos.
DE-AC06-76RLO1830, % PNNL
DE-SC0007983, % Wayne State
DE-SC0009824, % Florida
DE-SC0009973, % VPI
DE-SC0010073, % South Carolina
DE-SC0010118, % Carnegie Mellon
DE-SC0010504, % Hawaii
DE-SC0011784, % Cincinnati
DE-SC0012704; % BNL
%last group
and
%Vietnam
the National Foundation for Science and Technology Development (NAFOSTED) 
of Vietnam under contract No 103.99-2018.45.

\bibliography{conf.bib}

\end{document}